\documentclass[12pt,preprint]{aastex}

\usepackage{natbib}
\usepackage{lscape}
\usepackage[hyperindex,breaklinks]{hyperref}
\shorttitle{Infrared Spectroscopy of Halos of Edge-on Galaxies}
\shortauthors{Rand et al.}

\begin{document}


\title{Infrared Spectroscopy of the Diffuse Ionized Halos of Edge-on Galaxies}


\author{Richard J. Rand}
\affil{Department of Physics and Astronomy, University of New
Mexico, 800 Yale Blvd, NE, Albuquerque, NM 87131}
\email{rjr@phys.unm.edu}

\author{Kenneth Wood}
\affil{School of Physics and Astronomy, University of St. Andrews,
North Haugh, St. Andrews KY16 9SS, UK}
\email{kw25@st-andrews.ac.uk}

\author{Robert. A. Benjamin}
\affil{Department of Physics, University of Wisconsin at Whitewater,
800 West Main Street, Whitewater, WI 53190}
\email{ benjamir@uww.edu}

\author{Sharon E. Meidt}
\affil{Max-Planck-Institut für Astronomie, Königstuhl 17, D-69117 Heidelberg, Germany}
\email{meidt@mpia-hd.mpg.de}


\begin{abstract}

We present a study of ionized gas, PAHs, and H$_{\rm 2}$ emission in
the halos of three edge-on galaxies, NGC 891, NGC 5775 and NGC 3044,
based on $10-20$ $\mu$m {\it Spitzer Space Telescope} spectra.  The
[Ne$\,$III]/[Ne$\,$II] ratio, an excellent measure of radiation
hardness, rises with $z$ in the halo of NGC 891.  It is also higher in
the halo of NGC 5775 than in the disk.  NGC 3044 presents a more
confusing situation.  To explain the [Ne$\,$III]/[Ne$\,$II] as well as
optical line ratio behavior in NGC 891, we carry out a simple
exploration of parameter space with CLOUDY, which indicates a large
increase in radiation temperature with height.  Illustrative examples
of physical models using a Monte Carlo radiative transfer code show
that the rising neon ratio may be explained by adding a vertically
extended, hot stellar source to a thin disk of massive stars.
However, several other sources of hard spectra may be relevant.  PAH
features have scale heights of 430--530 pc in NGC 891 and 720--1080 pc
in NGC 5775, suggesting they can be transported by disk-halo flows.
Within NGC 891 and NGC 5775, scale heights are similar for all PAHs.
For NGC 891, the scale heights exceed that of 8 $\mu$m emission,
indicating a transition from more ionized to more neutral PAHs with
height.  Most PAH equivalent widths are higher in the halos.  H$_2$
17.03 $\mu$m emission with scale heights of $550-580$ pc in NGC 891
and 850 pc in NGC 5775 suggests a molecular component in a
surprisingly thick layer.

\end{abstract}





\section{Introduction}

Gaseous halos of spiral galaxies have grown in importance in recent
years as they are at the interface between galaxies and their
environments, and as such are the sites of many physical processes
that dictate galaxy evolution.  Supernovae and stellar winds in disks
can pump mass and energy into halos setting up a disk-halo cycle of
gas (\citealt{1989ApJ...345..372N}), depending on the level of star
formation activity.  Halos may also contain gas accreting onto
galaxies, either primordial or from companions
(\citealt{2008A&ARv..15..189S}).  Gas from these two sources may
interact dynamically, thermally, and chemically.  Such feedback from
star formation is believed to be important for understanding the growth of
galaxies and their current properties \citep{1991ApJ...379...52W}.

The discovery of thick layers of interstellar gas and dust, whether
referred to as ``halos'' or ``extraplanar'' components, has opened up
a window on these issues.  In the past twenty years or so, such thick
layers have been discovered in just about every component of the ISM,
especially in X-rays (e.g.,
\citealt{2006A&A...448...43T,2004ApJS..151..193S}), HI (e.g.,
\citealt*{2007AJ....134.1019O}), radio continuum (e.g.,
\citealt*{2006A&A...457..121D}), diffuse ionized gas (DIG; e.g.,
\citealt{2003A&A...406..505R,1996ApJ...462..712R}), and dust
(\citealt{1999AJ....117.2077H,1998ApJ...507L.125A,2000A&AS..145...83A}).
Particularly for the DIG, X-ray, radio continuum and dust components,
their brightness and extent correlate with the level of disk star
formation, both within and among galaxies, indicating an origin in a
disk-halo flow (\citealt{1996ApJ...462..712R,1998PASA...15..106R,
  2003A&A...406..493R,2006A&A...457..779T,2004ApJS..151..193S,2006A&A...457..121D,
  1999AJ....117.2077H}).  Yet, some halo gas may be due to continued
primordial infall onto galaxy disks, which some lines of evidence
indicate is necessary to maintain star formation (e.g.,
\citealt*{1997ApJ...477..765C}; \citealt*{2010ApJ...717..323B}).
Also, the metallicities of many Galactic High Velocity Clouds are as
low as 0.1 Z$\sun$ (\citealt{2004Ap&SS.289..381W}), possibly
indicating a mixing of primordial and processed gas.  Finally, lagging
DIG (e.g., \citealt{2007ApJ...663..933H}) and HI
(\citealt{2007AJ....134.1019O}) halos have also been interpreted as a
signature of such mixing (\citealt{2008MNRAS.386..935F}), although
other explanations are possible \citep{2002ASPC..276..201B}.

\subsection{Spectroscopy of DIG}

The DIG is a valuable tracer of energetic processes occurring in
gaseous halos, largely through the substantial leverage provided by
observations of emission lines.  Studies of halo or \lq\lq
extraplanar\rq\rq DIG in external edge-ons have been motivated by the
detailed characterization of the DIG (or Reynolds Layer) of the Milky
Way (\citealt{1990ApJ...349L..17R,2009RvMP...81..969H}), and the
questions that have arisen therefrom.  One of the most important
questions is what keeps the DIG ionized.

Despite the large vertical extent of extraplanar DIG layers in many
galaxies (e.g., \citealt{1996ApJ...462..712R}), there is much evidence
from various diagnostic line ratios that the primary source of
ionization is radiation leaking from a thin disk of massive stars
(e.g., \citealt*
{1997ApJ...474..129R,1998ApJ...501..137R,1997ApJ...483..666G,2001ApJ...551...57C,
  2002ApJ...572..823O,1999ApJ...523..223H,2003ApJ...586..902H,2006ApJ...652..401M}).
In the Milky Way, such radiation is also the only source that can meet
the energetic requirement of keeping the DIG layer ionized
(\citealt{1990ApJ...349L..17R}).  Most commonly observed is that the
ratios of [S$\,$II]$\lambda\lambda 6716,6731$ and
[N$\,$II]$\lambda\lambda 6548,6583$ to H$\alpha$ generally increase
with distance from the suspected ionizing source, in the Milky Way and
in external galaxy disks and halos (all emission lines, and ionization
energies necessary to create the responsible ions, relevant for this
paper are listed in Table 1).  This is a signature of a falling
ionization parameter, $U$, (e.g.,
\citealt{1994ApJ...428..647D,1998ApJ...501..137R}) but other factors
have been argued to affect line ratios significantly -- namely
elevated gas temperatures in the diffuse gas
(\citealt{2001ApJ...548L.221R}; \citealt{1999ApJ...523..223H}) and
hardening of the radiation field during its propagation
(\citealt{2003ApJ...586..902H,2004MNRAS.353.1126W}, hereafter WM).

The situation is further complicated by the spatial behavior of other
line ratios, such as [O$\,$III]$\lambda 5007$/H$\beta$,
[O$\,$I]$\lambda 6300$/H$\alpha$, and He$\,$I$\,\lambda 5876$/H$\alpha$
(see \citealt{2009RvMP...81..969H} for a summary).  The first of these
ratios, in particular, shows behavior in many galaxies suggesting that
some DIG ionization occurs through processes other than
photo-ionization by a thin disk of massive stars.  In several external spiral
galaxies, both edge-on and more face-on, this ratio increases or stays
relatively constant with distance from the ionizing stars (e.g.,
\citealt*{1998ApJ...501..137R,
  2000ApJ...537L..13R,1999AJ....118.2775G};
\citealt{2002ApJ...572..823O,
  2003ApJ...586..902H,2003ApJ...592...79M}).  An increase with height
has also been found in the Reynolds Layer in the inner Galaxy
(\citealt{2005ApJ...630..925M}).  This may be partially explained by
elevated gas temperatures, although \citet{2001ApJ...551...57C} find
this explanation insufficient for NGC 5775 and UGC 10288.  In cases such
as these and the well-studied NGC 891, other sources of ionization are
indicated, with shocks receiving the most attention (e.g.,
\citealt{2001ApJ...551...57C}).  Not all extraplanar DIG layers may
require such a source, however.  Some show a drop of
[O$\,$III]/H$\beta$ with height, $z$
(\citealt{2000A&A...362..119T,2003ApJ...592...79M}).  One of these is
NGC 3044, especially clear in the more westerly of the two long slit
spectra presented by \citet{2000A&A...362..119T}.

A further complication is the assumed abundances.  In general, solar
or ``ISM'' abundances are assumed in photo-ionization models
(e.g., \citealt{1994ApJ...428..647D,2000ApJ...528..310S}; WM), but one can
expect some sensitivity of line ratios to abundances, either directly,
or through variation in the cooling efficiency.  \citet{2000ApJ...528..310S}
find that most line ratios vary inversely with abundance in their
photo-ionization models, although \citet{1994ApJ...428..647D} and WM
find a more complicated relationship.  Given the possibility that halo
gas contains a mix of disk-halo cycled and primordial gas, abundance
variations may well be important.

Finally, one should mention the complication caused by extinction,
which has been found to be significant even at $z=1-2$ kpc from the
disk in several edge-on spirals, including NGC 891 (\citealt[][and
  references therein]{1999AJ....117.2077H,2000AJ....119..644H}).  We
crudely estimate in this paper that there should be many magnitudes of
visual extinction in the midplanes of the galaxies we report on.
Therefore, apart from the effect of extinction on line ratios, the
depth along the line of sight to which we are probing optically likely
varies significantly with $z$.

With these various difficulties in interpreting optical line ratios,
it is desirable to have a diagnostic of ionization with a more
straightforward interpretation.  Such an opportunity is provided by
the Infrared Spectrograph (IRS; \citealt{2004ApJS..154...18H}) on
board the {\it Spitzer Space Telescope}
(\citealt{2004ApJS..154....1W}), which allows measurement of the ratio
of the 15.56 $\mu$m [Ne$\,$III] (ionization potential 41.0 eV) and
12.81 $\mu$m [Ne$\,$II] (ionization potential 21.6 eV) lines, as well
as others.  This ratio provides a diagnostic of the hardness of
ionizing radiation that is relatively insensitive to extinction,
gas-phase abundances, and gas temperature (being low excitation lines
in warm gas) -- three of the biggest sources of confusion for the
optical lines [the ratio is enhanced in low-metallicity galaxies, but
  this is attributed to the associated harder stellar radiation fields
  (e.g.  \citealt{2010ApJ...712..164H},
  \citealt{2009ApJ...704.1159H})].  Pioneering work with the {\it
  Infrared Space Observatory} (ISO) first demonstrated the power of
this diagnostic (e.g., \citealt{2000ApJ...539..641T};
\citealt{2002ApJ...566..880G}; and \citealt{2003A&A...403..829V}).

In \citet*[hereafter RWB]{2008ApJ...680..263R} we presented our first
measurements of this ratio in the DIG halo of NGC 891 (assumed
distance 9.5 Mpc), at $z=1$ kpc on opposite sides of the disk.  The
ratio was found to be elevated in both extraplanar pointings relative
to the disk.

If the DIG is ionized by radiation leaking out of the disk, then the
same stars, with the same metallicities, are responsible for disk and
halo ionization, and our result for NGC 891 would therefore be due to
propagation effects.  To understand whether such a scenario is viable,
we combined the neon ratio with optical line ratios along a long slit
running through the IRS pointings, and modeled line emission from the
DIG layer using the 2D Monte-Carlo radiative transfer code of
\citet*{2004MNRAS.348.1337W}.  The code is also capable of 3D
simulations.  The models include the effects of radiation field
hardening and explore various temperatures and luminosities of the
radiation field, and were tailored to the diffuse ISM density
distribution of NGC 891.  As applied to optical line ratios for the
Reynolds layer and the halo of NGC 891, WM had found that spectral
hardening reduced the need for a non-ionizing heat source, yet it was
still not possible to match all the line ratio data, so that
additional heating and/or a secondary ionization source still seems to
be required.  With the neon ratio included, RWB found that no model
could reproduce the observed low values of this ratio and its rise
with $z$; neither could they predict a rising [O$\,$III]/H$\beta$, or
the typical values of [O$\,$I]/H$\alpha$, and He$\,$I/H$\alpha$.  They
are more successful in reproducing the rise of [S$\,$II]/H$\alpha$ and
[N$\,$II]/H$\alpha$, at least semi-quantitatively.  We emphasize that
including a non-ionizing heat source or a metallicity gradient should
affect the optical line ratios significantly but not the neon ratio.
These results are obviously problematic for such photo-ionization
models.

Clearly, one wishes to know whether the values of the neon ratio
measured by RWB, and their enhancement off the plane, are typical of
DIG layers, whether this infrared ratio reflects trends in optical
ratios, especially [O$\,$III]/H$\beta$ (the second ionization
potential of oxygen is 35.1 eV, nearly as high as that of neon), and
whether the neon ratio in general presents such difficulties for
photo-ionization models.  We have therefore extended our observations
to $z=2$ kpc on both sides of the plane in NGC 891 (where
[O$\,$III]/H$\beta$ rises more dramatically), and observed disk and
halo positions in two other edge-ons, NGC 5775 and NGC 3044 [we use
the same distances, 24.8 Mpc and 16.1 Mpc, respectively, as in
\citet{2000ApJ...536..645C} and \citet{1997ApJ...490..247L})], where
coincident optical spectra along slits perpendicular to the disk are also
available.  The former galaxy is more actively star forming than NGC 891
(\citealt{1996ApJ...462..712R}), while the long-slit spectroscopy
indicates a strongly rising [O$\,$III]/H$\alpha$ with $z$
(\citealt{2000ApJ...537L..13R}).  By contrast, as mentioned above, NGC
3044 shows a falling [O$\,$III]/H$\beta$, especially for one long slit
whose location is the focus of our IRS observations.  It also has a
relatively bright extraplanar DIG layer (\citealt{2003A&A...406..493R}).

We reconsider photo-ionization models in light of our results,
focusing on NGC 891.  We begin with a general exploration of
parameter space using the CLOUDY code \citep{1998PASP..110..761F},
asking the simple question of whether there is any combination of
ionization parameter, radiation temperature and gas
temperature that can reproduce the infrared and optical line ratios in
NGC 891 independently at each height.  That exploration opens up a
wide range of parameter space with a new set of constraints to
consider in physical models, and suggests a tractable way to explore
multiple ionization sources in future work.  Consistent with these
findings, we present an initial exploration of this space with new
Monte Carlo radiative transfer models that include a second,
vertically extended stellar ionizing source.

\subsection{Dusty Halos}

The dust content of gaseous halos can provide clues to their origin.
There is plenty of evidence for dusty halos through both emission
(e.g., \citealt{2009MNRAS.395...97W}) and absorption (e.g.,
\citealt{1999AJ....117.2077H}).  That dusty halos are powered by star
formation driven disk-halo cycling of gas is suggested by the
correlation among galaxies between the presence of extraplanar
H$\alpha$ extinction and emission (e.g.,
\citealt{1999AJ....117.2077H}).  It seems most likely that dust is
transported upwards with the gas.  However, other explanations for
halo dust are possible (see \citealt{1997AJ....114.2463H} for a
general discussion), including radiative acceleration of grains
(\citealt{1991ApJ...366..443F}), which may lift clouds to heights of a
few hundred pc through gas-dust coupling.  It has also been argued
that grain lifetimes are much shorter than the typical residence time
in the ISM (\citealt{2003ARA&A..41..241D}), implying much dust
formation and modification in the ISM.  It seems more likely that such
dust formation would take place in the denser disk environment than in
halos, although the energetic processes involved in disk-halo flows
may modify dust properties.  This latter point may be important for
understanding dust evolution and provides motivation for studies such
as ours.

The IRS provides the ability to study emission
from Polycyclic Aromatic Hydrocarbons (PAHs;
\citealt*{1984A&A...137L...5L,1985ApJ...290L..25A,2001ApJ...556..501B};
\citealt{2001ApJ...560..261B}; \citealt*{2007ApJ...657..810D}) - 
molecules up to about 20$\AA$ in size containing up to a few thousand
C atoms which in SINGS galaxies are found to contribute a few percent
of the dust mass (\citealt{2007ApJ...663..866D}).  Emission features in the
$10-20$ $\mu$m range are thought to arise from more neutral PAHs
relative to features in the $6-9$ $\mu$m range (\citealt{2007ApJ...657..810D};
\citealt{2008ApJ...679..310G}).

It is uncertain how PAHs relate to the larger dust grains, but it has
become clear in recent years that PAH emission is suppressed relative
to the emission from larger grains in the immediate vicinity of
recently formed stars (e.g., \citealt{2007ApJ...665..390L}), a result
attributed to destruction of PAH molecules by UV radiation (e.g.,
\citealt{2008ApJ...682..336G}).  A key piece of evidence for this
picture is the decrease of PAH Equivalent Widths (EW) with increasing
radiation field hardness (at least in the $20-40$ eV range), as
measured by [Ne$\,$III]/[Ne$\,$II], in star forming regions
(\citealt{2006A&A...446..877M}; \citealt{2006ApJ...639..157W};
\citealt{2008ApJ...678..804E}), although there seems to be no
correlation below [Ne$\,$III]/[Ne$\,$II]\ $\sim$ 1
(\citealt{2006ApJ...653.1129B}; \citealt{2008ApJ...682..336G}), the
regime that will be relevant here.

In halos, radiation fields are much weaker, whatever their hardness,
and it is less likely that processing of grains by radiation will be
important.  Furthermore, it should be pointed out that in the general
ISM as well as in halos, radiation field intensities are low enough so
that even the continuum in the $10-20$ $\mu$m range should be dominated
by grains heated by single photons rather than ones in thermal
equilibrium, according to \citet{2007ApJ...657..810D}.  We should
therefore expect that the PAH spectrum should not vary significantly
with radiation hardness or intensity for the kinds of environments
relevant here.  Rather, relative variations in PAH emission are likely
to reflect real changes in the PAH population, and would point to
other energetic processes likely to be relevant for halos, such as the
aforementioned photo-levitation, or shocks, where PAHs may be produced
(by grain collisions or sputtering; \citealt*{1996ApJ...469..740J}),
as well as destroyed \citep*{2010A&A...510A..36M}.

Imaging is beginning to reveal extended halos of PAHs in edge-on
galaxies (e.g., \citealt{2009MNRAS.395...97W} for NGC 891,
\citealt{2009MNRAS.396.1875H} for NGC 5775).  Spectroscopy reveals
more detailed information but fewer studies exist.  In RWB our limited
spectra provided a first look at scale heights and EWs of PAH features
in NGC 891.  Elsewhere, the halo of the starburst M82 has also been
studied spectroscopically, first by \citet{2006ApJ...642L.127E}.  More
recently, \citet{2008ApJ...679..310G} find that the intensity ratios
of both the 6.2 and 7.7 $\mu$m features relative to the 11.2 $\mu$m
feature decrease in this halo, which they interpret as a drop in the
contribution from ionized PAHs.  Otherwise, they find the mix of PAHs
in M82 is quite invariant.  In the same galaxy,
\citet{2008ApJ...676..304B} find higher EWs of neutral PAHs, but not
ionized PAHs, in the halo than in the disk, but they attribute this to
a drop in the continuum longward of 10 $\mu$m relative to the PAH
strength.

With our larger set of IRS spectra we can start to examine these
issues in the halos of more normal, albeit still quite actively star
forming, edge-on galaxies, and do so in more depth than was possible
in RWB.  Scale heights of PAH emission can be compared with those of
other vertically extended components to relate PAH and gaseous halos.
Scale heights of various features and disk-halo contrasts of their
EWs can be compared to search for modification of the PAH population with
height.

\section{Observations}

The data were taken on 2008 March 26 (NGC 5775), September 9 (NGC
891), and 2009 January 7--8 (NGC 3044) (program ID40284; PI: R. Rand)
using the staring mode of the IRS Short-High (SH) module on board {\it
  Spitzer}.  A log of the observations is given in Table 2.  The SH
module is a cross-dispersed echelle spectrograph providing spectral
coverage from 9.9 to 19.6 $\mu$m with a resolving power of $\sim$
600.  The aperture has dimensions of $4.7\arcsec$x$11.3\arcsec$.  In
staring mode the SH module nods between two pointings centered 1/3 and
2/3 of the way along the slit.

Coordinates of all pointings are given in Table 2, while Figures 1, 2,
and 3 show the locations of the apertures (including the NGC 891
apertures from RWB) on H$\alpha$ images of the three galaxies, as well
as the overlapping long slits discussed above.  For NGC 891, the new
observations were made at locations in the halo of NGC 891 centered at
a height of approximately $z= \pm 2$ kpc (hereafter ``east halo'' and
``west halo'' pointings) on opposite sides of a location in the disk
at 100$\arcsec$ from the center of the galaxy.  The reconstructed
pointings indicate that the actual slit centers were at heights of 2.0
and 1.9 kpc on the east and west sides, respectively.  The NGC 5775
halo pointing is at $z=2$ kpc on the SW side of the halo.  Three
locations in the disk below were observed (since a pointing NW of the
disk position directly underlying the halo position would include the
nucleus and thus perhaps not be representative of the disk, both
additional disk pointings are toward the SE).  For NGC 3044, our two
halo pointings are at heights of 0.7 and 1.2 kpc from the disk on the
NE and SW sides, respectively, and three disk locations were also
observed.

Because of the bright infrared background, separate sky observations
are necessary for IRS exposures of faint emission, particularly when
the emission is expected to fill the slit.  These were performed
contiguously with the target observations, using an exposure time
equal to that of the halo observations for each galaxy.

\clearpage
\begin{figure}
\epsscale{.80}
\includegraphics[scale=1,viewport=50 0 574 468]{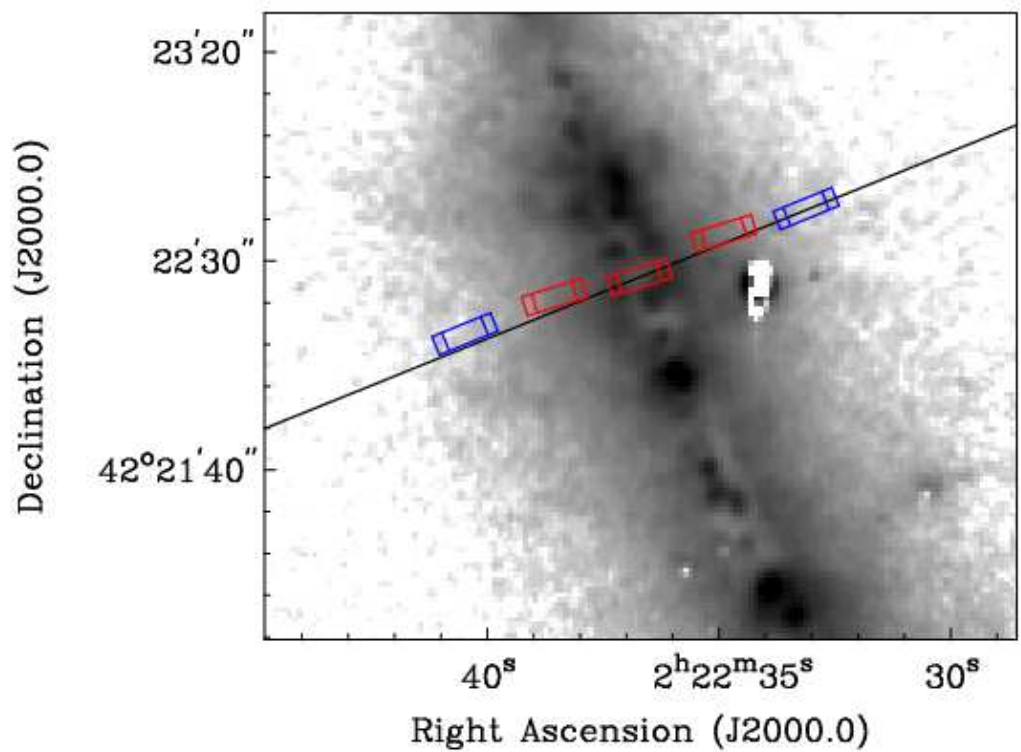}
\caption{Section of H$\alpha$ image of NGC 891
  (\citealt{1990ApJ...352L...1R}).  Boxes show IRS pointings,
  including in red the three pointings from RWB closest to the plane
  (for each pointing two overlapping boxes showing the two nods are
  drawn).  The solid line shows the orientation of the slit for the
  optical emission line data discussed.}
\label{fig1}
\end{figure}
\clearpage

\clearpage
\begin{figure}
\epsscale{.80}
\includegraphics[scale=1,viewport=50 0 574 468]{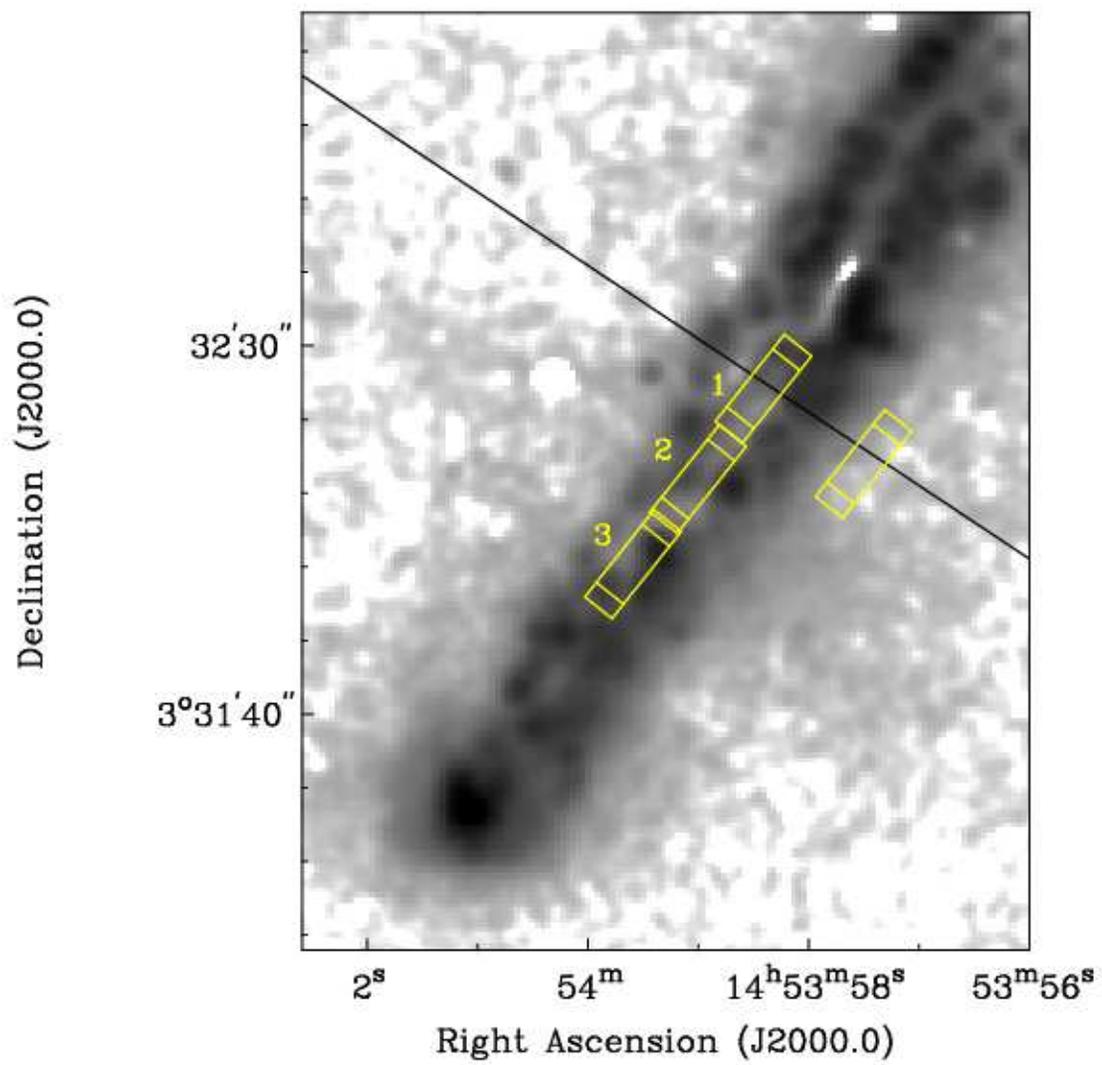}
\caption{IRS pointings for NGC 5775 overlaid on the H$\alpha$ image from
  \citet{2000ApJ...536..645C}.}
\label{fig2}
\end{figure}
\clearpage

\clearpage
\begin{figure}
\epsscale{.80}
\includegraphics[scale=1,viewport=50 0 574 468]{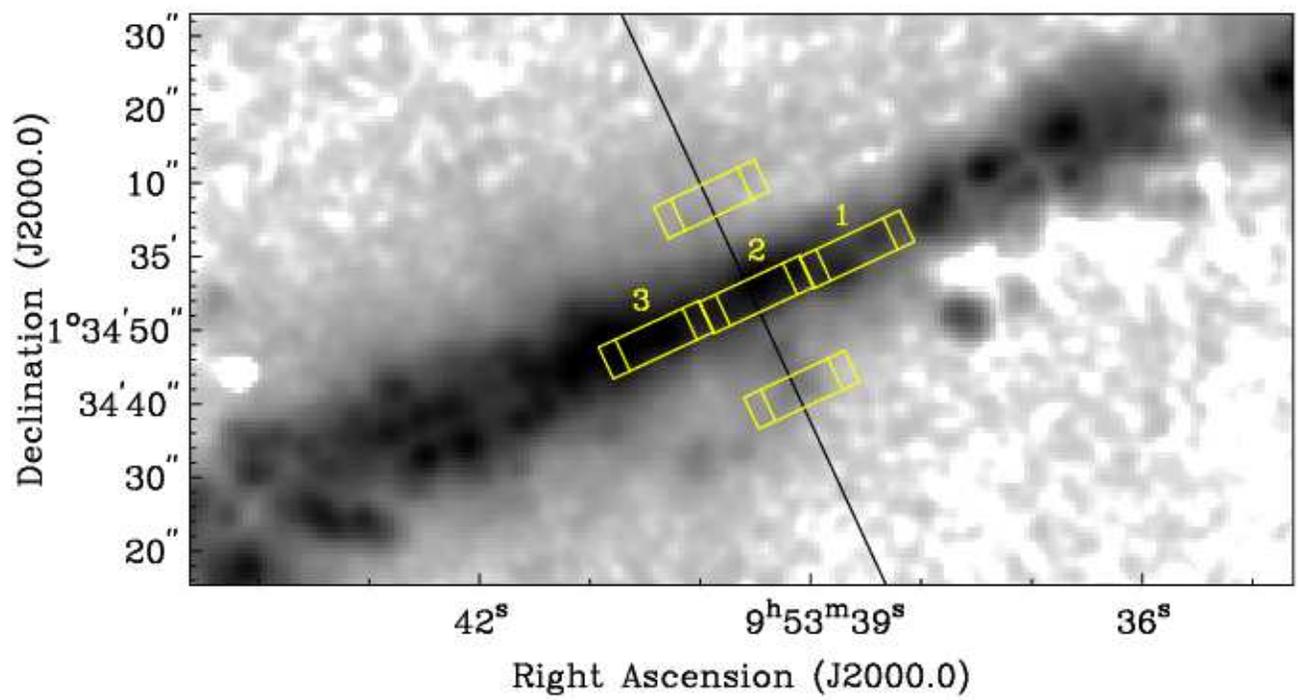}
\caption{IRS pointings for NGC 3044 overlaid on the H$\alpha$ image from
\citet{2000ApJ...536..645C}.}
\label{fig3}
\end{figure}
\clearpage

The data were processed through versions S17.2 (NGC 5775), S18.1 (NGC
891) and S18.5 (NGC 3044) of the IRS reduction pipeline, providing the
Basic Calibrated Data (BCD) products.  Post-BCD processing was carried
out according to the Infrared Spectrograph Data Handbook Version 2.0.
Charge accumulation can be present in long integrations, evidenced by
steady increases in the signal, but no significant effect was found in
our data.  ``Rogue'' pixels were cleaned using the contributed
IRSCLEAN\_MASK software.  For each of the AORs listed in Table 2 and
for each nod, the 2-D spectra were averaged using sigma-clipping at
the 3$\sigma$ level.  For each galaxy, the resulting sky spectra for
both nods were averaged and subtracted from all target observations.
All remaining extreme pixels not removed by IRSCLEAN\_MASK were
replaced with averages of adjacent pixels in the sky-subtracted
spectra.

One-dimensional spectra were extracted (and calibrated based on
standard star observations) with the package SPICE (version 2.2),
using the full aperture width.  Note that, due to slit loss correction
uncertainties, the flux calibration is only fully accurate if the
source is uniformly bright and covers an area much larger than the
slit.  In practice, our sources usually show some gradient across the
slit (at worst a factor of 3 for the NGC 3044 halo pointings; the halo
pointings of NGC 891 also show a significant gradient at about the 30--50\%
level).
However, for the most part we can argue that the flux corrections are
either minimal or accounted for in the scatter of our derived values.
The two halo pointings for NGC 891 yield similar fluxes, and our main
conclusions hold regardless of the scatter between these pointings
(the same is true of the data in RWB).  For NGC 5775, the three disk
pointings also show similar fluxes, especially for the PAH features,
which have a typical dispersion of 5\% from pointing to pointing.
This value may serve as a reasonable upper limit for slit loss
correction variations given the contribution from real structure in
the disk and other sources of error.  We therefore assume fluxes from
our halo pointing are also not significantly affected by slit loss
uncertainties.  For NGC 3044, apart from such uncertainties, our
results are unfortunately compromised by what we assume is a large
degree of real variation in the disk pointings.

A final spectrum for each pointing was formed by combining the
resulting 1-D spectra for each AOR and nod.  As the spectral orders
overlap, the ends of each order were removed from the final spectra.
The spectra are shown in Figures 4, 5, and 6.

\clearpage
\begin{figure}
\epsscale{0.8}
\includegraphics[angle=0,scale=1.3,viewport=0 100 374 468]{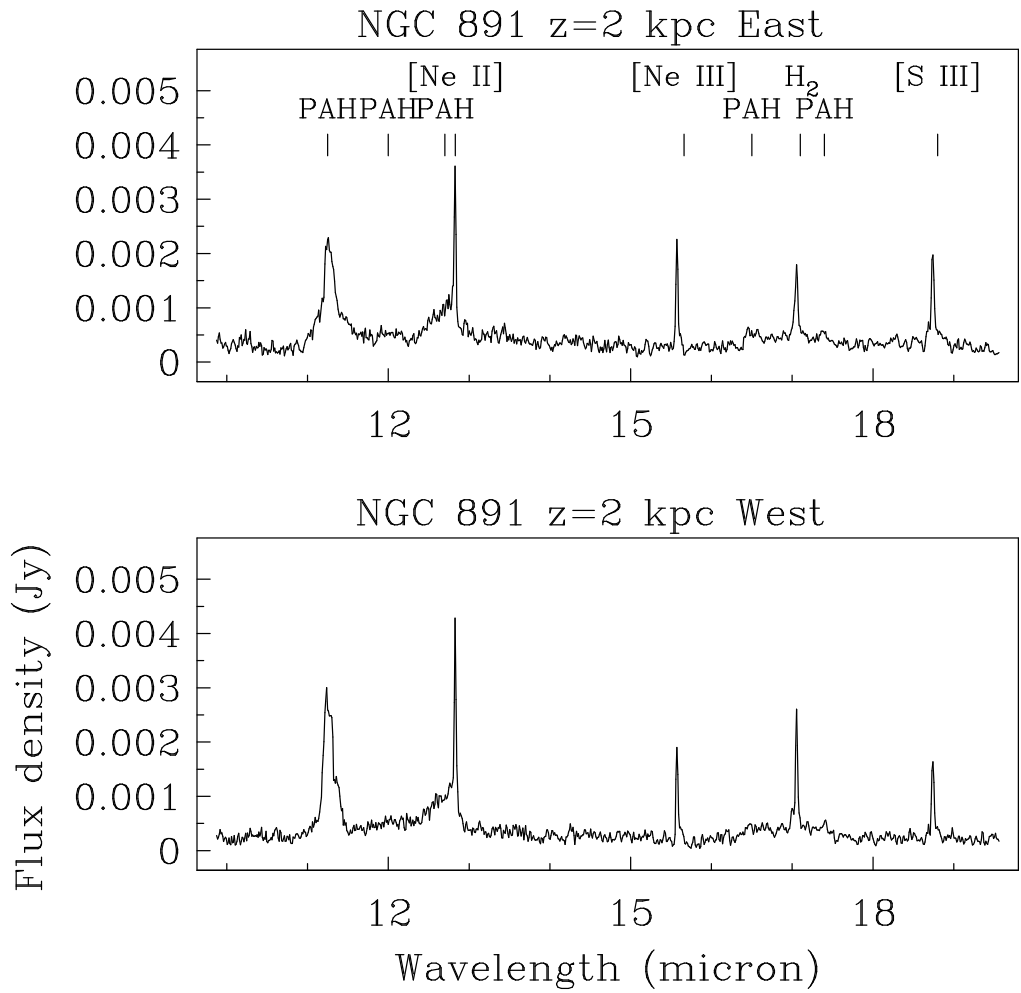}
\caption{IRS SH spectra of the east and west halo pointings of NGC 891
  at $z=2$ kpc.  Detected gas and the five strongest PAH features are
  indicated in the top panel.}
\label{fig4}
\end{figure}
\clearpage

\clearpage
\begin{figure}
\epsscale{.80}
\includegraphics[angle=0,scale=1,viewport=50 100 374 668]{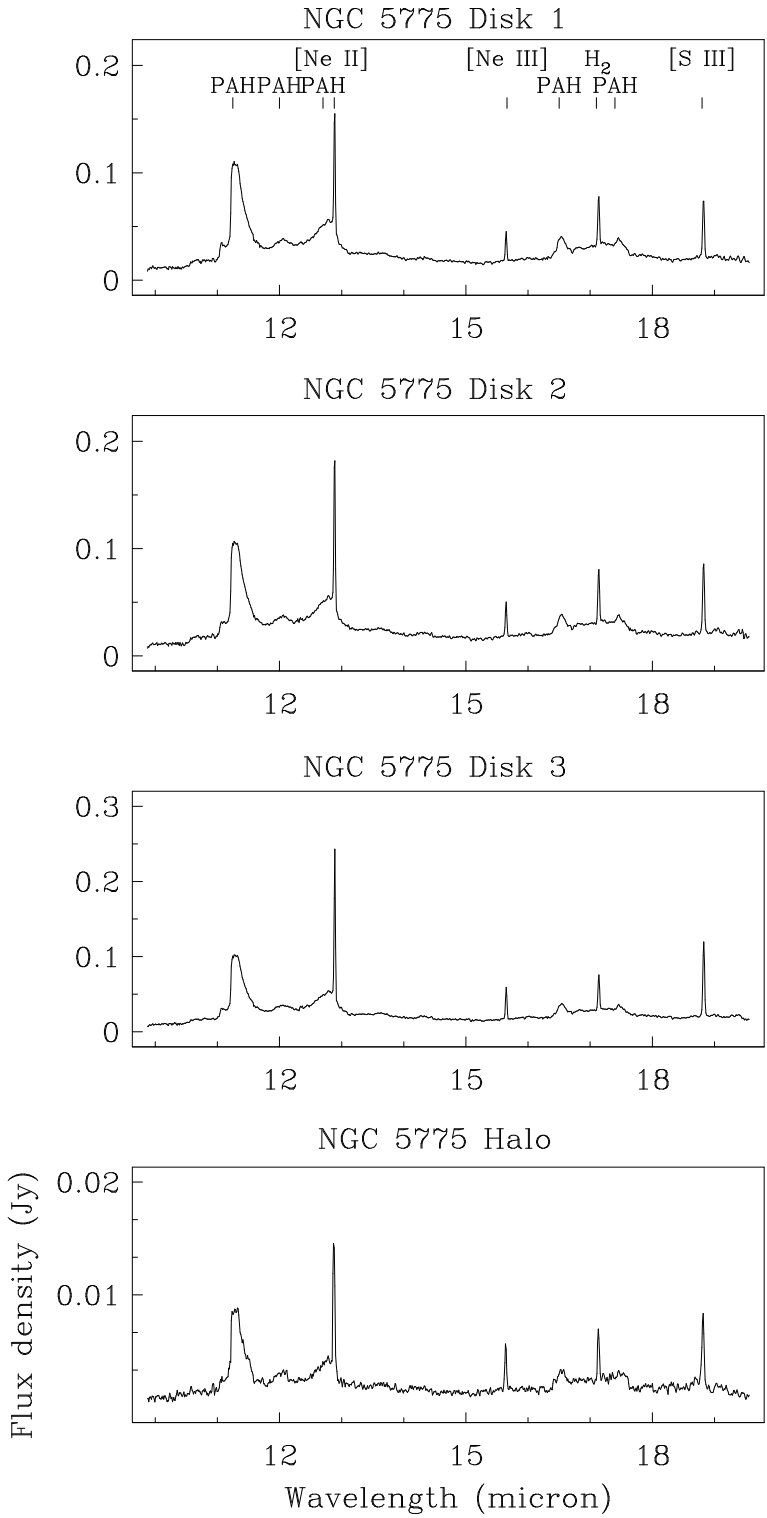}
\caption{Same as Figure 4 but for the disk and halo pointings of
  NGC 5775.}
\label{fig5}
\end{figure}
\clearpage

\clearpage
\begin{figure}
\epsscale{.80}
\includegraphics[angle=0,scale=.8,viewport=0 100 574 768]{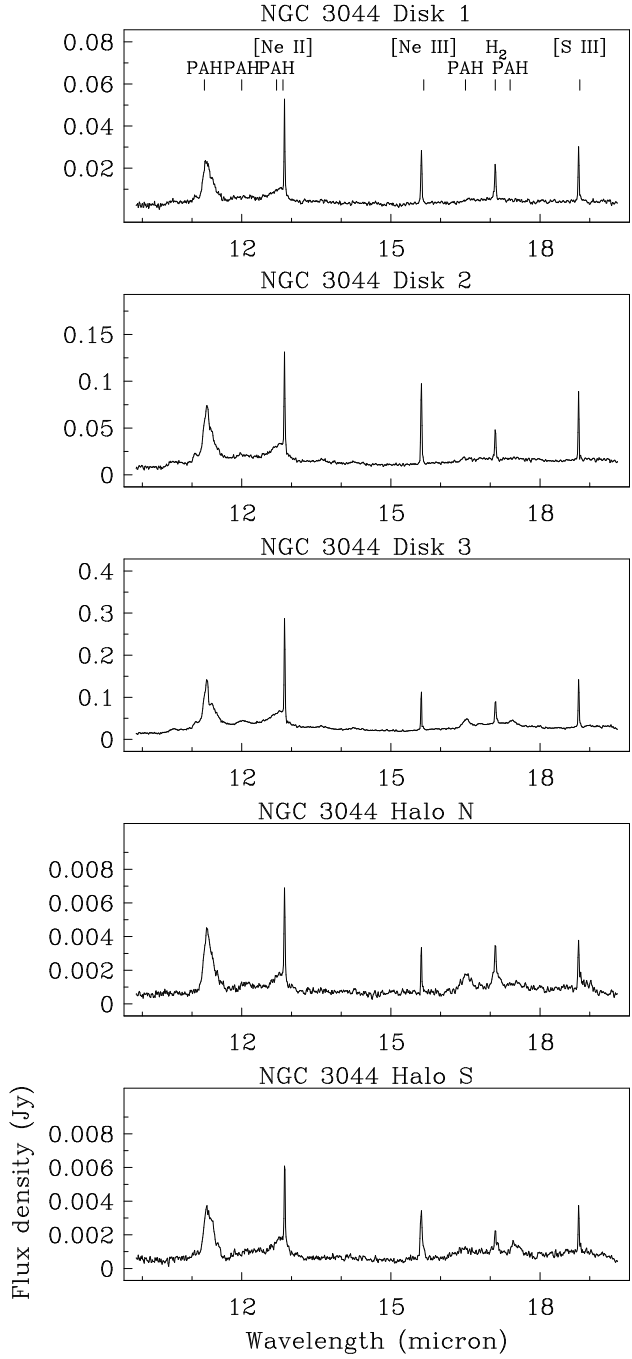}
\caption{Same as Figure 4 but for the disk and halo pointings of
  NGC 3044.}
\label{fig6}
\end{figure}
\clearpage

As in RWB, we checked that the fluxes in our halo pointings cannot be due to disk
emission convolved with the PSF of the IRS SH module using the {\it Spitzer}
contributed software STINYTIM.  In all cases, such a contribution from
disk emission to the halo fluxes is negligible compared to the measured values.

Line intensities and EWs were measured with the IRAF\footnote[1]{IRAF
  is distributed by the National Optical Astronomy Observatory, which
  is operated by the Association of Universities for Research in
  Astronomy, Inc., under cooperative agreement with the National
  Science Foundation.} program {\it splot} by summing pixels over the
line extent and subtracting a linear baseline.  We compute error bars
on intensities and EWs following RWB, taking into account noise in the
spectra and the estimated uncertainties in the flux scale of IRS data,
as described in the Infrared Spectrograph Data Handbook Version 3.1.
Random noises measured around the relatively flat continuum at
15 $\mu$m wavelength range from about 1 mJy for the disk pointings to
0.07 mJy for the NGC 891 halo pointings.

We identify the following stronger PAH features and give in
parentheses the wavelength range used to estimate their fluxes and and
to define a linear continuum for the EWs: 11.2 $\mu$m ($11.0-11.8$
$\mu$m), 12.0 $\mu$m ($11.9-12.2$ $\mu$m), 12.7 $\mu$m ($12.4-13.2$ 
$\mu$m), 16.5 $\mu$m ($16.3-16.8$ $\mu$m), and 17.4 $\mu$m ($17.3-17.7$
 $\mu$m).  Weaker features are also seen in some cases at 13.6, 14.3 and
15.9 $\mu$m.  All of these features are commonly seen in galaxies
(e.g., \citealt{2007ApJ...656..770S}).  The spectra from RWB were also
reanalyzed to ensure that all features were measured in a consistent
way.

Defining the continuum for these PAH features is difficult and in some
cases the ``continuum'' values may include blended emission from
discrete features.  It may therefore be instructive to consider an EW
which minimizes this possible contamination.  Hence, we also consider
EWs of the combined 11.2, 12.0 and 12.7 $\mu$m and weaker 13.6 and
14.3 $\mu$m complexes using a linear continuum from 10.8 to 14.5
$\mu$m, and the combined 17 $\mu$m complex using a linear continuum
from 16.3 to 17.7 $\mu$m.  Spectral decompositions of low-resolution
IRS spectra suggest that the emission in these two wavelength ranges
is dominated by discrete PAH features (e.g.,
\citealt{2007ApJ...656..770S}), while the ranges used for continuum
are indeed essentially free of discrete features.  Even for this type
of EW, however, there is concern that the 10.8 $\mu$m continuum may be
lowered by silicate absorption centered at 9.7 $\mu$m (e.g.,
\citealt{2006ApJ...637..774C}).  We will discuss the effects of
extinction in general on our results below.  Just as importantly, even
the ``continuum'' in the SH band is probably due to single-photon PAH
heating: \citet{2007ApJ...663..866D} conclude this to be the case for
16 $\mu$m emission observed in SINGS galaxies with the ``blue
peak-up'' array of the IRS.

\section{Results}

\subsection{Ionized Gas Phase Emission Lines}

We detect [Ne$\,$II] 12.81 $\mu$m, [Ne$\,$III] 15.56 $\mu$m, and
[S$\,$III] 18.71 $\mu$m in all pointings.  Line intensities and ratios
are presented in Tables $3-9$, along with upper limits on [S$\,$IV]
10.51 $\mu$m emission.

The data for NGC 891 also warrant a graphical presentation.
[Ne$\,$III], [Ne$\,$II] and [S$\,$III] intensities for NGC 891 from
the current data and RWB are plotted in Figure 7a while Figure 7b
shows [Ne$\,$III]/[Ne$\,$II] and [S$\,$III]/[Ne$\,$II] ratios.  The
falloff of intensity with $z$ is not well described by an exponential
for some lines.  Nevertheless, we report equivalent exponential
emission scale heights in Table 10 and plot them in Figure 8 (error
bars reflect the scatter in the data on the two sides of the disk) to
give a sense of the vertical extent of the emission.  The key result
is that the trend of rising [Ne$\,$III]/[Ne$\,$II] for this galaxy
found by RWB is continued by the $z=2$ kpc pointings.  The near
constancy of [S$\,$III]/[Ne$\,$II] reported by RWB, however, is only
continued in the east halo pointing at $z=2$ kpc; the value in the
west halo is significantly higher.  This is somewhat surprising given
the similar ionization potentials of S++ and Ne+.

For NGC 5775, there is a small but significant (at about the 5$\sigma$
level) [Ne$\,$III]/[Ne$\,$II] contrast between the three (remarkably
consistent) disk and one halo pointings.  The equivalent exponential
emission scale heights for the three lines are listed in Table 10 and
plotted in Figure 8 (error bars reflect the scatter in the disk
values).  For NGC 3044, the disk and halo pointings show a tremendous
amount of scatter for the intensities and both line ratios, precluding
any trends from being seen.

\clearpage
\begin{figure}
\epsscale{.80}
\includegraphics[scale=.8,viewport=0 0 574 668]{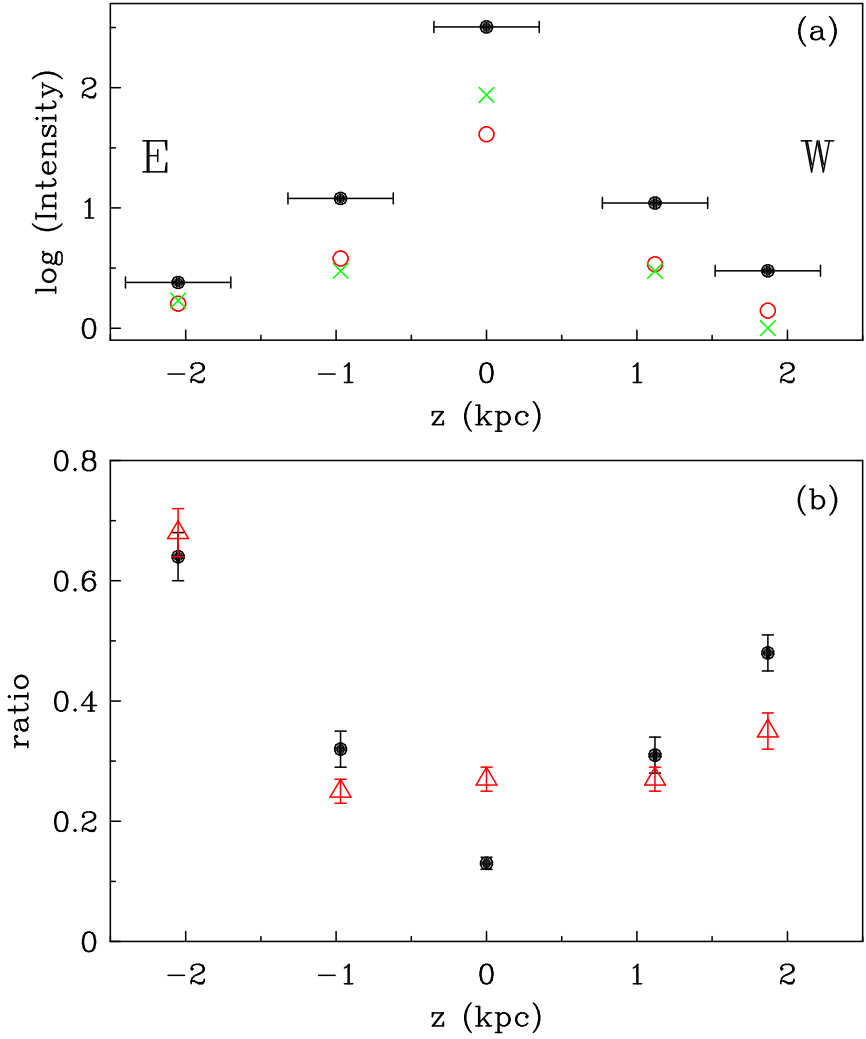}
\caption{({\it a}) The log of intensity (in units of 10$^{-17}$ erg
  cm$^{-2}$ s$^{-1}$ arcsec$^{-2}$) vs. $z$ for the [Ne$\,$III] (red
  open circles), [Ne$\,$II] (black filled circles) and [S$\,$III]
  (green crosses) lines in NGC 891.  Vertical error bars are smaller
  than the symbol sizes and are omitted for clarity but listed in
  Table 3. ({\it b}) [Ne$\,$III]/[Ne$\,$II] (filled black circles) and
  [S$\,$III]/[Ne$\,$II] (open red triangles) vs. $z$ in NGC 891.
  Horizontal error bars on the [Ne$\,$II] points in ({\it a}) indicate the
  slit extent in the direction perpendicular to the disk. }
\label{fig7}
\end{figure}
\clearpage

\clearpage
\begin{figure}
\epsscale{.80}
\includegraphics[scale=.8,viewport=0 0 574 668]{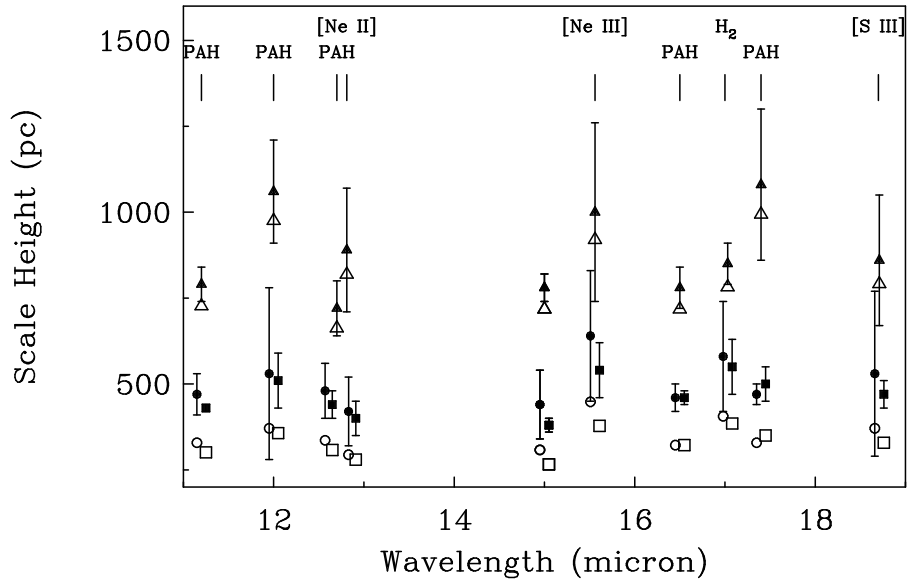}
\caption{({\it a}) Scale heights of gas emission lines and PAH
  features in the east (circles) and west (squares) halos of NGC 891
  and the halo of NGC 5775 (triangles), ordered by the wavelength of
  the feature.  Solid symbols represent values uncorrected for
  extinction and include error bars, while open symbols include a
  rough extinction correction (see text). The NGC 891 points have been
  offset slightly in wavelength for clarity.}
\label{fig8}
\end{figure}
\clearpage


As in RWB, we crudely estimate the effects of extinction at the
positions of our midplane pointings on the above results based on CO
and HI observations and assuming Galactic extinction properties.
There are many assumptions in this calculation, including the
conversion from CO intensity to H$_2$ column density, the gas-to-dust
ratio, and the near-IR extinction law.  Resolution mismatches between
these maps and the IRS SH observations are also a source of error.
For NGC 891, we use the same gas column densities as in RWB.  For NGC
5775, we use SEST CO 1--0 and uniformly-weighted VLA HI maps from
\citet{2001A&A...377..759L}, with the caveat that the CO map has a
relatively low angular resolution of 43'' (the resolution of the HI
map is 13.5'').  For NGC 3044, we use the uniformly weighted VLA HI
map from \citet{1997ApJ...490..247L} at about 21'' resolution, and
(kindly provided to us by J. Irwin) the CO 2--1 map from
\citet{1998PhDT.........4L} at 21'' resolution, along with an average
CO 2--1/1--0 ratio of 0.41.  We convert CO intensities to H$_2$ column
densities assuming a conversion factor of $3 \times 10^{20}$ mol
cm$^{-2}$ (K km s$^{-1}$)$^{-1}$, as used by
\citet{1993ApJ...404L..59S} for NGC 891.  With these assumptions, we
estimate total gas column densities at the position of our central
disk pointing in each galaxy.

From these column densities, we estimate the midplane extinction at a
representative wavelength of 15 $\mu$m, making the same assumptions as
in RWB: the Galactic relation between gas column density and $A_V$
(\citealt*{1978ApJ...224..132B}), the standard Galactic extinction law
to relate $A_V$ to $A_K$ (e.g., \citealt{2003ARA&A..41..241D}), and an
approximate relation $A_{15}/A_K \approx 0.4$ from Milky Way
observations by \citet{2000A&A...356..705R} and
\citet{2006A&A...446..551J} [although the model of
  \citet{2001ApJ...548..296W} for $R_V = 3.1$ indicates a value closer
  to 0.2].  We also assume that the dust and emitting gas are well
mixed and thus only half of the dust lies in front of the gas on
average.  For NGC 891, we estimate a midplane extinction at 15 $\mu$m
of 0.5 mag, which supercedes the value reported by RWB.  For NGC 5775
we find a value of only 0.12 mag, and for NGC 3044 the extinction is
negligible.  Since these values are subject to such large
uncertainties, they can be little more than illustrative.  For the
extinction curve of \citet{2000A&A...356..705R}, extinction should be
about 50\% higher at 17 $\mu$m, and perhaps 100\% higher at $11-12$
$\mu$m, due to silicate absorption at 9.7 and 18 $\mu$m.  These
midplane extinctions are also likely to be lower limits given the low
resolution of many of the CO and HI observations.  It may well be that
the bulk of the molecular and atomic gas is in unresolved layers for
many of these maps.  However, given the lack of observations of
sufficiently high resolution, we do not attempt to quantify this
possibility here by making any assumptions.

A midplane extinction of 0.5 mag, more appropriate for NGC 891,
reduces scale heights by 30\%, while a value of 0.1 mag, more
appropriate for NGC 5775, reduces scale heights by 8\%.  For the
aforementioned infrared extinction curves, extinction for the
[Ne$\,$II] and [Ne$\,$III] lines should be comparable and therefore
line ratios should not be signficantly affected.  Extinction for the
[S$\,$III] line may be somewhat higher.

\subsection{PAH Emission Features}

Tables 3, 5, 7 and 8 show the intensities and EWs of PAH features for the
three galaxies, as well as the intensity integrated over the most
feature-free spectral range, $14.5-15.5$ $\mu$m, which we refer to as
the 15 $\micron$ continuum.  Intensities for the 11.2 $\mu$m feature
include the secondary peak at about 11.0 $\mu$m.  EWs of the combined
features discussed in \S 2 are also listed, and are hereafter referred
to as EW$_{\rm short}$ and EW$_{\rm long}$.  For the NGC 891 pointings
in RWB, EW$_{\rm short} = 3.64 \pm 0.11, 3.75 \pm 0.11$ and $3.83 \pm
0.11$, EW$_{\rm long} = 0.56 \pm 0.02, 0.78 \pm 0.02$ and $0.73 \pm
0.02$, and the 15 $\mu$m continuum intensities are $800 \pm 61, 32 \pm
2$, and $35 \pm 3$ (all in units of 10$^{-17}$
erg~cm$^{-2}$~s$^{-1}$~arcsec$^2$) for the disk, east halo at $z=1$ kpc
and west halo at $z=1$ kpc, respectively.  The spectra allow us
to characterize the vertical distribution of PAH and continuum
emission and to search for general differences of PAH emission
properties between disks and halos.

The rather large scatter in PAH feature brightness in the disk
pointings of NGC 3044 preclude meaningful statements about the
vertical distribution of emission.  Hence we focus on NGC 891 (where
the GTO spectra discussed by RWB confirm that our disk pointing is
representative) and NGC 5775, where the three disk pointings show a
very small scatter in PAH emission.  Figure 9 shows PAH and 15 $\mu$m
continuum intensities vs. $z$ on a logarithmic scale for NGC 891.
Most features can reasonably be described by exponential
distributions; and results of exponential fits to the east and west
sides separately are presented in Table 10 and Figure 8.  Scale
heights range from 430 to 530 pc for PAH features (with an average of
$475 \pm 30$ pc, the error accounting for the dispersion of the ten
values in Table 10), similar to the range reported in RWB, while the
15 $\mu$m continuum has scale heights of 440 pc and 380 pc for the
east and west sides, respectively.  Hence, we have a hint that PAH
emission is more vertically extended than the continuum emission,
although the uncertainties are large.  The EW analysis below will
strengthen this conclusion.  Table 10 and Figure 8 also show scale
heights for NGC 5775, assuming an exponential distribution.  For
these, intensities for the three disk positions have been averaged,
and the error bars on the scale heights reflect the scatter and the
average uncertainty of the three measurements.  Here, the scale
heights are much larger than in NGC 891, ranging from 720 to 1080 pc.
But within each galaxy, PAH scale heights are similar, given the error
bars.

\clearpage
\begin{figure}
\epsscale{.80}
\includegraphics[scale=.8,viewport=0 0 574 668]{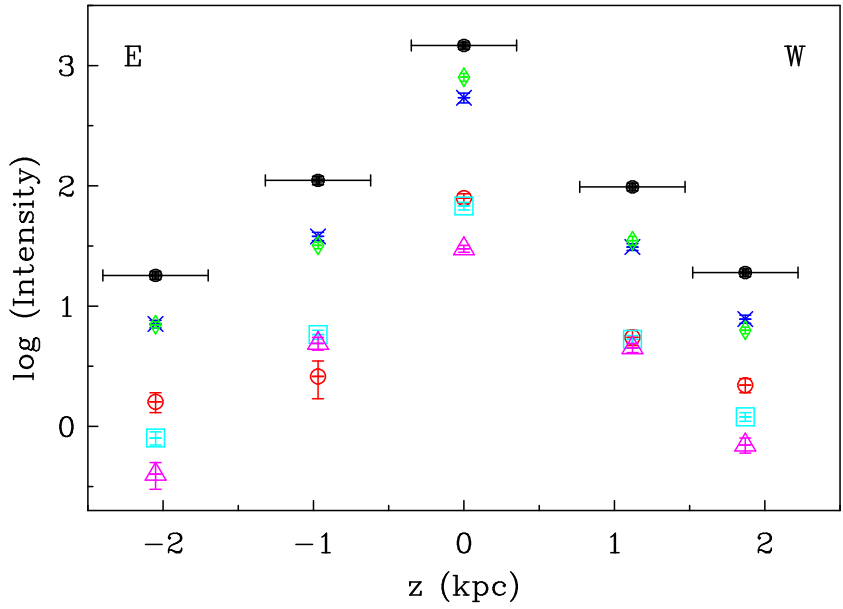}
\caption{The log of intensity (in units of 10$^{-17}$ erg cm$^{-2}$ s$^{-1}$
  arcsec$^{-2}$) vs. $z$ for the 11.2 (filled black circles),
  12.0 (open red circles), 12.7 (blue crosses), 16.5 (cyan squares)
  and 17.4 $\mu$m (magenta triangles) PAH features and the 15 $\mu$m
  continuum (green diamonds) in NGC 891.  Horizontal error bars on the
  11.2 $\mu$m PAH points indicate the slit width.}
\label{fig9}
\end{figure}
\clearpage

Extinction will lower scale heights as described above for the
gas lines.  Scale heights of the 11.2 and 17.4 $\mu$m features are
likely to be reduced more by extinction than those of the other features.

The PAH emission scale heights in NGC 891 with no extinction
correction are larger than that of the CO-emitting gas layer, which
has a FWHM of 225 pc (\citet{1993ApJ...404L..59S}).  The HI layer is
modeled by \citet*{2007AJ....134.1019O} as thin and thick exponential
layers with scale heights increasing with radius from $<0.3$ to 0.5
kpc for the former and 1.25 to 2.5 kpc for the latter, which contains
30\% of the gas.  Hence, the PAH emission scale heights are clearly
larger than the CO scale height and are perhaps more comparable to
that of the thin HI layer, with or without our extinction correction.
For NGC 5775, \citet{1994ApJ...429..618I} finds an exponential scale
height of 1.1 kpc for the HI, while the molecular layer thickness is
presumably substantially less but is not well known.  So again, it is
probably the case that the PAH scale heights are comparable to or
somewhat less than that of the HI layer.

More significant results can be extracted from the EWs.  These are not
sensitive to any possible extinction contrast between disk and halo,
although individual values may be affected by the shape of the
extinction curve.  Also, for NGC 3044, while intensities
show much scatter among the disk positions, the EWs are more
consistent.  Figure 10 shows EWs of the five bright PAH
features, and EW$_{\rm short}$ and EW$_{\rm long}$, plotted as a
function of $z$ for NGC 891, NGC 5775 and NGC 3044, respectively.
These figures make it clear that, for most features (all five in
Figure 10a, three of five in Figures 10c and 10e, and all in Figures 10b,
10d and 10f), EWs are higher in the halo than in the disk, no matter
which way they are measured.  Note that for NGC 891 this conclusion
differs from that in RWB, where a significant disk-halo EW contrast
was found only for the 17.4 $\mu$m feature.  We attribute this
difference to the more careful EW measurements in this work.  Given
the scatter among galaxies, no feature stands out here as having a
significantly higher disk-halo EW contrast than others.

\clearpage
\begin{figure}
\epsscale{.80}
\includegraphics[scale=0.85,viewport=30 0 574 628]{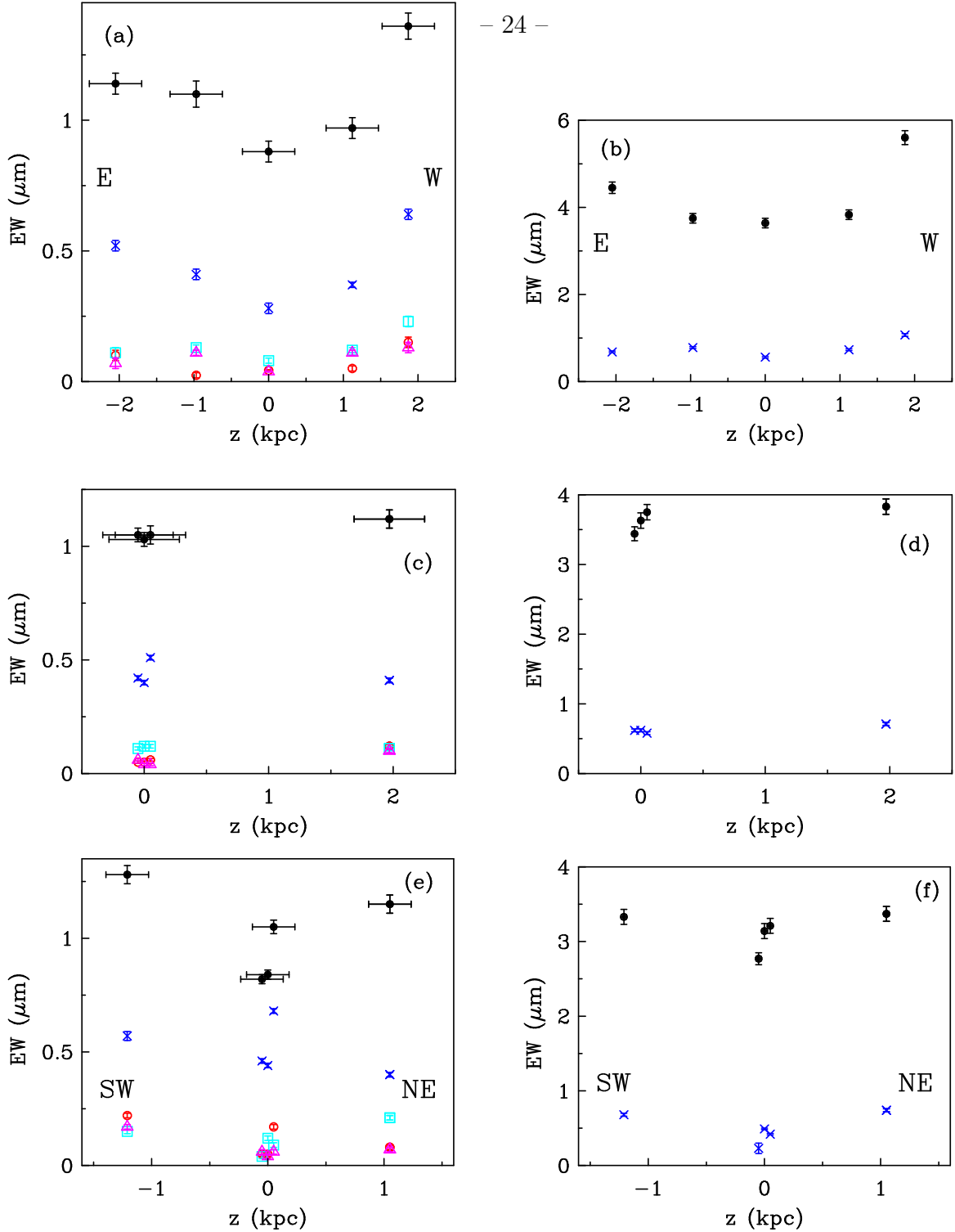}
\caption{Equivalent Width vs. $z$ for PAH features.  ({\it a}) 11.2
  (filled black circles), 12.0 (open red circles), 12.7 (blue
  crosses), 16.5 (cyan squares) and 17.4 $\mu$m (magenta triangles)
  features in NGC 891.  ({\it b}) EW$_{\rm short}$ (filled black
  circles) and EW$_{\rm long}$ (blue crosses) in NGC 891.  ({\it c})
  and ({\it d}) same as ({\it a}) and ({\it b}), respectively but for
  NGC 5775.  ({\it e}) and ({\it f }) same as ({\it a}) and ({\it b}),
  respectively, but for NGC 3044.  Horizontal error bars on the 11.2
  $\mu$m PAH points indicate the slit extent in the direction
  perpendicular to the disk.}
\label{fig10}
\end{figure}
\clearpage



\subsection{H$_{\rm 2}$ Emission}

The H$_2$ {\it S}(1) $J=3-1$ 17.03 $\mu$m line is detected in all
pointings.  H$_2$ mid-IR emission lines arise from rotational
excitation in warm gas.  Such emission is usually attributed to PDRs
(e.g., \citealt{2004ARA&A..42..119V}) but may also arise in shocks
(e.g., \citealt{2006ApJ...649..816N}).  Scale heights for best-fit
exponentials are given in Table 10 and plotted in Figure 8.  For NGC
891, the values of 550 and 580 pc are somewhat higher than found by
RWB without the $z=2$ kpc data points.  Again, the scale height in NGC
5775, 850 pc, is significantly larger.  Extinction (see above) may
lower these values somewhat.  Like the PAH features, these scale
heights are comparable to the HI scale heights in both galaxies.  It
is difficult to interpret these results in terms of column densities
of infrared-emitting molecular hydrogen from only one spectral line.
A full analysis would require several rotational transitions to be
observed so that excitation temperatures and ortho-para ratios could
be modeled (e.g., \citealt{2008ApJ...675..316B};
\citealt{2007ApJ...669..959R}; \citealt{2006ApJ...649..816N}).
Nevertheless, the discovery of emission from H$_2$ with scale heights
in the range $500-800$ pc suggests a molecular gas component with a
surprisingly large vertical extent.  We do note, however, that there
have been indications of a faint, vertically extended CO emitting
component in NGC 891 (\citealt{1992A&A...266...21G};
\citealt{1993PASJ...45..139S}).  Again the greater scale height in NGC
5775 suggests that disk-halo flows are somehow responsible for this
vertical extent.

\section{DIG Photo-ionization modeling}

As stated in \S 1, the [Ne$\,$III]/[Ne$\,$II] ratio provides an
excellent diagnostic of ionization as it is relatively free of effects
of extinction, gas temperature and gas abundances that can complicate
interpretation of common optical line ratios.  The extraplanar
pointings at $z=2$ kpc on either side of the plane of NGC 891 continue
the trend of rising [Ne$\,$III]/[Ne$\,$II] with $z$ found by RWB.
There is a significant but much smaller rise in NGC 5775, while
scatter in the ratio prevents any trend from being observed in NGC
3044.  All values of the ratio in DIG range from about 0.1 to 1.  A
similar trend is seen in M82 by \citet{2008ApJ...676..304B}, with
values increasing from about 0.15 in the disk to a high of 0.27 a few
arcseconds above the disk.

These results indicate a higher ionization state in the halos of these
two galaxies, confirming what had been suggested by the more
problematic [O$\,$III] line (\citealt{2001ApJ...551...57C}).  For what
follows, we focus on NGC 891, where we have by far the most
information on halo line ratios.  In addition to
[Ne$\,$III]/[Ne$\,$II] values, Figure 11 shows the dependence with height of
[N$\,$II]$\lambda 6583$/H$\alpha$, [S$\,$II]$\lambda 6716$/H$\alpha$,
[O$\,$I]$\lambda 6300$/H$\alpha$, [O$\,$III]$\lambda 5007$/H$\beta$,
and He$\,$I $\lambda 5876$/H$\alpha$ (a relatively clean optical
diagnostic of ionization, for which extinction is the major concern
and the faintness of the helium line the major challenge) along a slit
running through nearly all of our IRS pointings (as shown in Figure
1).  The optical data are from \citet{1997ApJ...474..129R} and
\citet{1998ApJ...501..137R}.

\clearpage
\begin{figure}
\epsscale{.80}
\includegraphics[scale=.8,viewport=0 100 574 768]{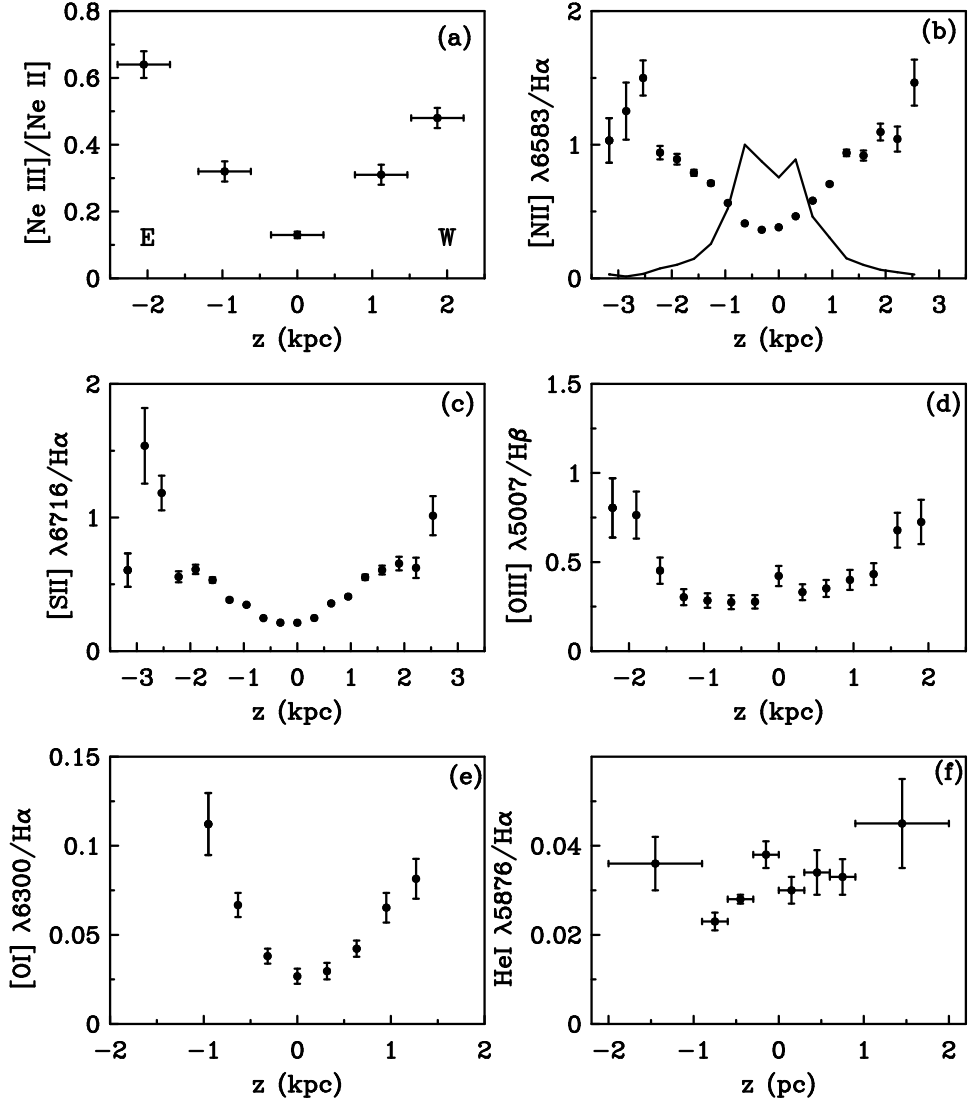}
\caption{Dependence of (a) [Ne$\,$III]/[Ne$\,$II], (b)
  [N$\,$II]$\lambda 6583$/H$\alpha$, (c) [S$\,$II]$\lambda
  6716$/H$\alpha$, (d) [O$\,$I]$\lambda 6300$/H$\alpha$, (e)
  [O$\,$III]$\lambda 5007$/H$\beta$, and (f) He$\,$I $\lambda
  5876$/H$\alpha$ on $z$ for the NGC 891 long slit spectrum discussed
  in the text.  The H$\alpha$ profile, normalized to unit intensity,
  from \citet{1998ApJ...501..137R}, is shown in (b).  Horizontal error
  bars in (a) and (f) reflect the extent over which the data have been
  averaged.  In the other panels, the intensities were averaged over
  317 pc.  The data are from \citet{1997ApJ...474..129R} and
  \citet{1998ApJ...501..137R}.}
\label{fig11}
\end{figure}
\clearpage

We wish to understand what sources of ionization can explain these
behaviors, with particular attention paid to the constraints provided
by the neon ratio.  

Our main focus in this paper is on stellar photo-ionization, but by
necessity we attempt to broaden the parameter space heretofore
considered in ionization models in general.  We take two approaches.
We begin by using the CLOUDY code \citep{1998PASP..110..761F} to ask,
independently at each height of 0, 1 and 2 kpc, whether there are any
combinations of radiation temperature, ionization parameter,
and gas temperature that can reproduce the observed ratios (\S 4.1).
By simply exploring this parameter space, we temporarily set aside the
issue of whether any solution corresponds to a reasonable physical or
``structural'' model - that is, we do not try to explain why there
might be changes in radiation temperature with $z$, or what balance of
heating and cooling may lead to a certain gas temperature.  However,
the difficulty encountered in previous modeling, even before the
strong constraints provided by the infrared lines were available,
suggests that this naive approach is a useful way to constrain what
kinds of physical models may be viable.

Before turning to physical models, we make several remarks about
shocks (\S 4.2).  Although standard shock models in combination with
photo-ionization have been explored to explain the optical line ratio
behavior in the DIG of several edge-ons (see \S I), we point out
the difficulty in applying such models to halo environments, and that
our approach offers a more feasible way of considering them.

In considering secondary sources of ionization, we have been partly
motivated by the possibility of a second, hot, vertically extended
source of photo-ionization as a way of producing a rising
[Ne$\,$III]/[Ne$\,$II] ratio with $z$.  Indeed, our CLOUDY results
will give us further motivation to consider such a source.  As an
illustration of such a physical model, in our second approach (\S
4.3) we will then turn to the Monte Carlo photo-ionization code
described in \citet{ 2004MNRAS.353.1126W} and employed in RWB, adding
such a component to the thin layer of massive stars we considered in
that paper.  This code is well suited to exploration of additional
stellar sources.  Our CLOUDY results will open up a broad range of
parameter space to consider in such models, and it is beyond the scope
of this paper to explore fully this parameter space - hence we
restrict ourselves to a few illustrative cases.

\subsection{Line Ratio Constraints on Physical Conditions vs. Height in 
NGC 891} 

The increasing level of ionization in the gaseous halos indicates an
increase in the ionization parameter, the hardness of the radiation
field, the gas temperature, or a combination of all three. One
difficulty in explaining the observed emission lines lies in the fact
that {\it at any given height} there is a range of densities. There is
also quite probably a range of temperatures, either due to variation
in photoionization heating or the presence of other forms of
dissipational heating, such as shocks, turbulence, cosmic rays, or magnetic
reconnection, that are expected to become important at low
densities. It is therefore to be expected that it will be difficult to
match the observed ratios with a single model.  Uniform density models
seem likely to fail, and models with {\it a priori} density variations
would have to guess the right combinations of density variations to
match all the line ratios. Past attempts at photoionization modeling
have confirmed that it is challenging to match the observed emission
lines.

\begin{figure}
\begin{center}
\includegraphics[scale=0.7, angle=90]{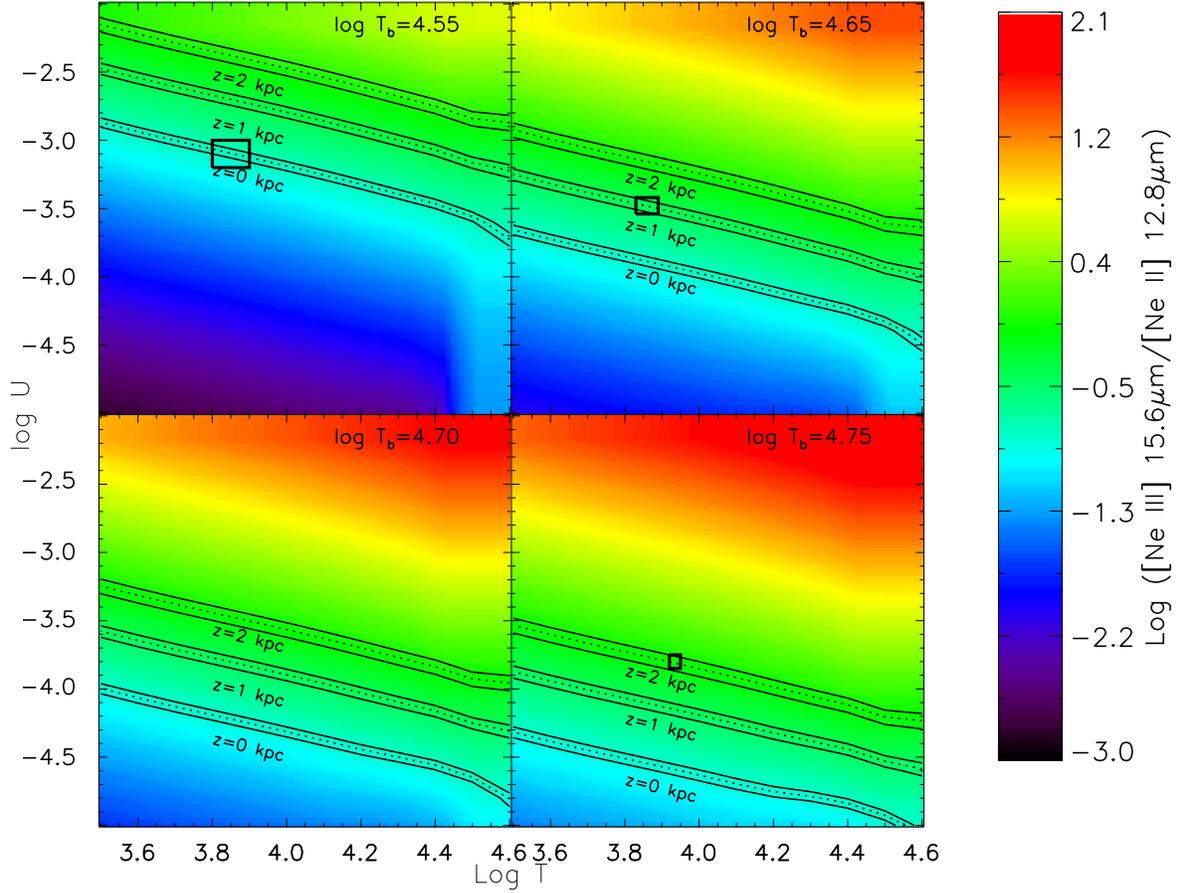}
\caption{Predicted emission line ratios for [Ne$\,$III]/[Ne$\,$II] as
  a function of gas temperature, $\log T$, and ionization parameter,
  $\log U$, for four values of the blackbody temperature $\log
  T_{b}=$4.55, 4.65, 4.70, and 4.75. (Note that the $T_{b}$ \lq\lq
  slices\rq\rq\ shown in each of four quadrants will differ in
  subsequent plots.)  The measured ratios at $z=0$, 1, and 2 kpc are
  indicated as diagonal dashed lines, with the solid bands indicating
  the range allowed when uncertainties are included. The $z=0$ kpc and
  $z=2$ kpc sheets are separated by 1.25 dex in $\log T$, 0.7 dex in
  $\log U$, and 0.1 dex in $\log T_{b}$. The allowed parameter space
  range after incorporating selected other line ratios is shown (by
  black boxes) for convenience and discussed further in the text and
  subsequent figures. }
\end{center}
\label{zones}
\end{figure}

With several measured line ratios, particularly in the halo of NGC
891, this problem is overdetermined. In this subsection, we use these
line ratios to try diagnose the {\it local} physical conditions in the
gas as a function of height, {\it independent of any model to produce
  these physical conditions}. In an ideal situation, in which one
density/temperature phase dominates the observed emission, one set of
physical parameters would match all the line ratios simultaneously.
Although some line ratios remain problematic, we find that four key
line ratios can be matched simultaneously and yield a change of
physical conditions versus height that is physically plausible.

We have used the photoionization code CLOUDY v8.0 for a solar
metallicity gas to calculate how measured line ratios in the gaseous
halo of NGC 891 vary as a function of three parameters: the ionization
parameter, $U=\Phi_{H}/(cn_{H})$, the \lq\lq radiation
temperature\rq\rq\ (for convenience we parametrize the spectral shape
with a blackbody spectrum), $T_{b}$ , and the gas temperature,
$T$. The ionization parameter is defined in terms of the
plane-parallel flux of photons with energy greater than 13.6 eV,
$\Phi_{H}$, the speed of light, $c$, and the particle density of
hydrogen atoms, $n_{H}$. The ionization parameter and blackbody
temperature determine the photoionization rate of the different ions
by specifying the flux and spectral shape of the radiation field.  An
absorbed stellar flux or spectra from X-ray emitting gas could also be
used.

Figure 12 shows how the [Ne$\,$III]/[Ne$\,$II] ratio varies as a function of
these three parameters.  In the range of parameters space that we
considered, this ratio spans five orders of magnitude!  With the
relatively small uncertainties in the measured line ratio at each of
the three heights, a thin sheet of the three-dimensional parameter
space is allowed. To go from the ratio observed at $z=0$ kpc to that
observed at $z=2$ kpc, holding two of the three parameters fixed,
would require a whopping 1.25 dex ($\times 18$) increase in the gas
temperature, a 0.7 dex ($\times 5$) increase in ionization parameter,
or a modest 0.1 dex ($\times 1.25$) increase in blackbody
temperature. The relative insensitivity of this line ratio to gas
temperature, which only occurs because of the temperature dependence
of the recombination rates, makes it a key diagnostic.

\begin{figure}
\begin{center}
\includegraphics[scale=0.7, angle=90]{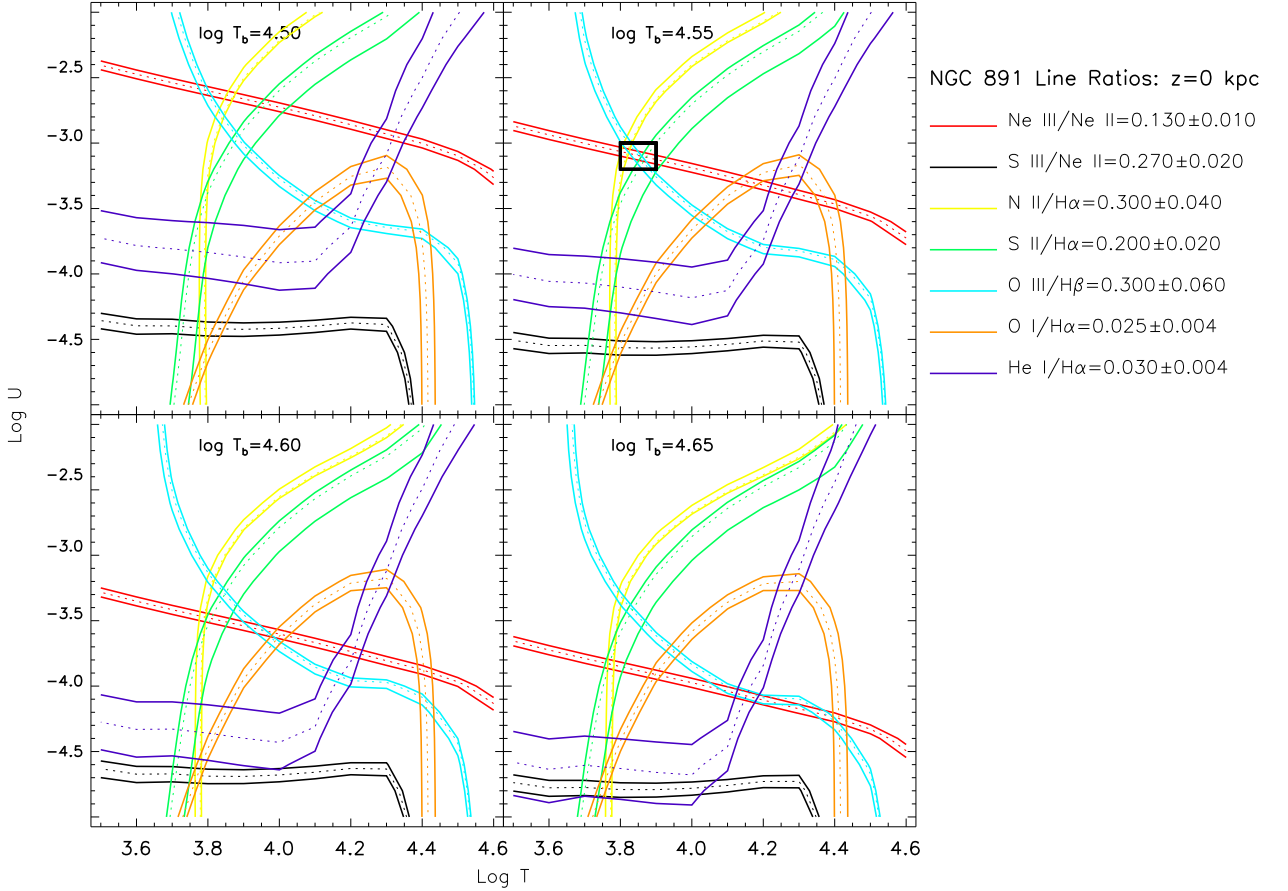}
\caption{Allowed parameter space ($\log T$ and $\log U$) for four
  values of the blackbody temperature $\log T_{b}=$4.50, 4.55, 4.60,
  and 4.65. The line ratios used to constrain the parameter space are
  indicated by the dotted lines, with uncertainty ranges shown with
  the solid lines. Each line ratio is indicated by a different color
  with the observed value (and uncertainty) given to the right of the
  plot (for doublets, the same line is plotted as in Figure 11).  A
  successful model would have all the bands (or in the three
  dimensional parameter space, sheets) intersect in a single locus. In
  this and subsequent plots, the line ratios [Ne$\,$III]/[Ne$\,$ II],
  [S$\,$II]/H$\alpha$, [N$\,$II]/H$\alpha$, and [O$\,$III]/H$\beta$
  have a common intersection point noted with a black box.
  Uncertainties of 1$\sigma$ on the line ratios have been used to
  define the uncertainties on the parameters.  The parameters for this
  intersection point are given in Table 11. }
\end{center}
\label{z0}
\end{figure}

Incorporation of the other observed gas line ratios produces equivalent
slices through parameter space, most of which have a stronger gas
temperature dependence.  The allowed parameter space based on measured
line ratios at $z=0$, 1, and 2 kpc are shown in Figures 13-15.  The
ratios [S$\,$II]/H$\alpha$ and [N$\,$II]/H$\alpha$ behave similarly,
requiring a higher gas temperature to match the observed data as the
ionization parameter increases.  [O$\,$III]/H$\beta$, on the other
hand, has the opposite behavior, requiring lower gas temperature as the
ionization parameter increases. The intersection point of these three
ratios already identifies a very limited section of parameter
space. With the addition of the infrared [Ne$\,$III]/[Ne$\,$II] ratio,
the problem becomes over-determined. The fact that the parameter space
allowed by the intersection point of these optical line ratios (as
well as He$\,$I/H$\alpha$ at $z=1$ kpc, although the uncertainties are
relatively large) also agrees with the [Ne$\,$III]/[Ne$\,$II] is
suggestive that one type of gas, with a unique set of physical
conditions, dominates the observed emission. (We will address the
remaining ratios shortly.)

\begin{figure}
\begin{center}
\includegraphics[scale=0.7, angle=90]{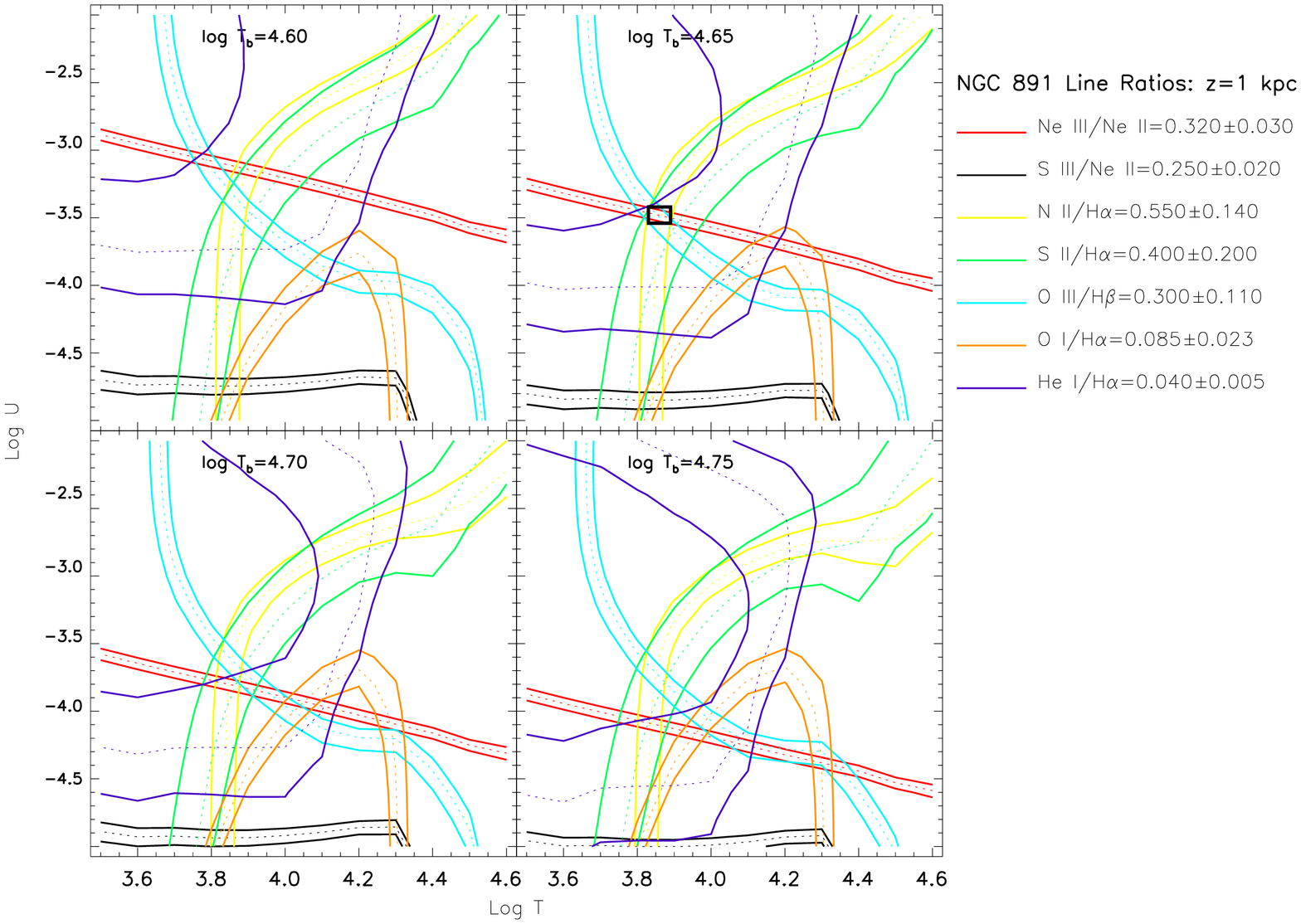}
\caption{Same as Figure 13 but for the observed NGC 891 line ratios at
  $z=1$ kpc, and for $\log T_{b}=$4.60, 4.65, 4.70, and 4.75.}
\end{center}
\label{z1}
\end{figure}

Although the intersection of these four line ratios in parameter space
is not perfect for $z=0$ kpc, it is satisfactory given the
uncertainties in the abundances for NGC 891. And for $z=1$ and 2 kpc,
it is possible to identify a unique set of parameters that
simultaneously match all four line ratios (and He$\,$I/H$\alpha$ at
$z=1$ kpc).  The derived parameters, as a function of height, are
given in Table 11 (the 1$\sigma$ uncertainties define a volume
in parameter space -- the quoted uncertainties are those projected onto
each axis).  This table also gives the derived value for the
(plane-parallel) flux of hydrogen ionizing photons,
$\Phi_{H}=cn_{H}U$, normalized to a density of $10^{-1}~{\rm
  cm^{-3}}$.  Since these values are derived solely from the line
ratios, it is possible that non-physical values could have been
obtained, and yet they seem plausible. First they indicate a {\it
  drop} in ionization parameter by a factor of five over a distance of
2 kpc (despite the increasing level of ionization).  When converted to
a flux of hydrogen ionizing photons, it indicates an incident flux at
$z=0$ kpc of $\Phi_{H}=n_{-1}(2.40 \pm 0.55) \times 10^{6}~{\rm
  photons~s^{-1}~cm^{-2}}$, where $n_{-1}= n_{H}/0.1 {\rm
  cm}^{-3}$. If we assume a constant density, only 41\% and 20\% of
the incident flux reaches $z=1$ kpc and 2 kpc, respectively.  Using a
representative gas density scale height of $h_{z}=2.3$ kpc (see RWB)
and a fixed filling factor, these decrease to 27\% and 8\% of the
incident flux at $z=1$ kpc and 2 kpc, respectively. Both the total
photon flux and the order of magnitude attenuation are comparable to
what is observed in the Galactic warm ionized medium, which has a
total flux of hydrogen-ionizing photons of $\Phi_{H}=2.6 \times
10^{6}~{\rm photons~s^{-1}~cm^{-2}}$ at $z=0$ kpc
\citep{1990ApJ...349L..17R} and $\Phi_{H} \sim 0.2 \times 10^{6}~{\rm
  photons~s^{-1}~cm^{-2}}$ well above the plane, based on the
measurement of H$\alpha$ emission from high velocity clouds
\citep{2002ApJ...572L.153T}. The values obtained for the Galaxy were
obtained by measuring the H$\alpha$ flux and not from any line ratio
information. It is encouraging that two different techniques yield
such similar numbers for two galaxies.

\begin{figure}
\begin{center}
\includegraphics[scale=0.7, angle=90]{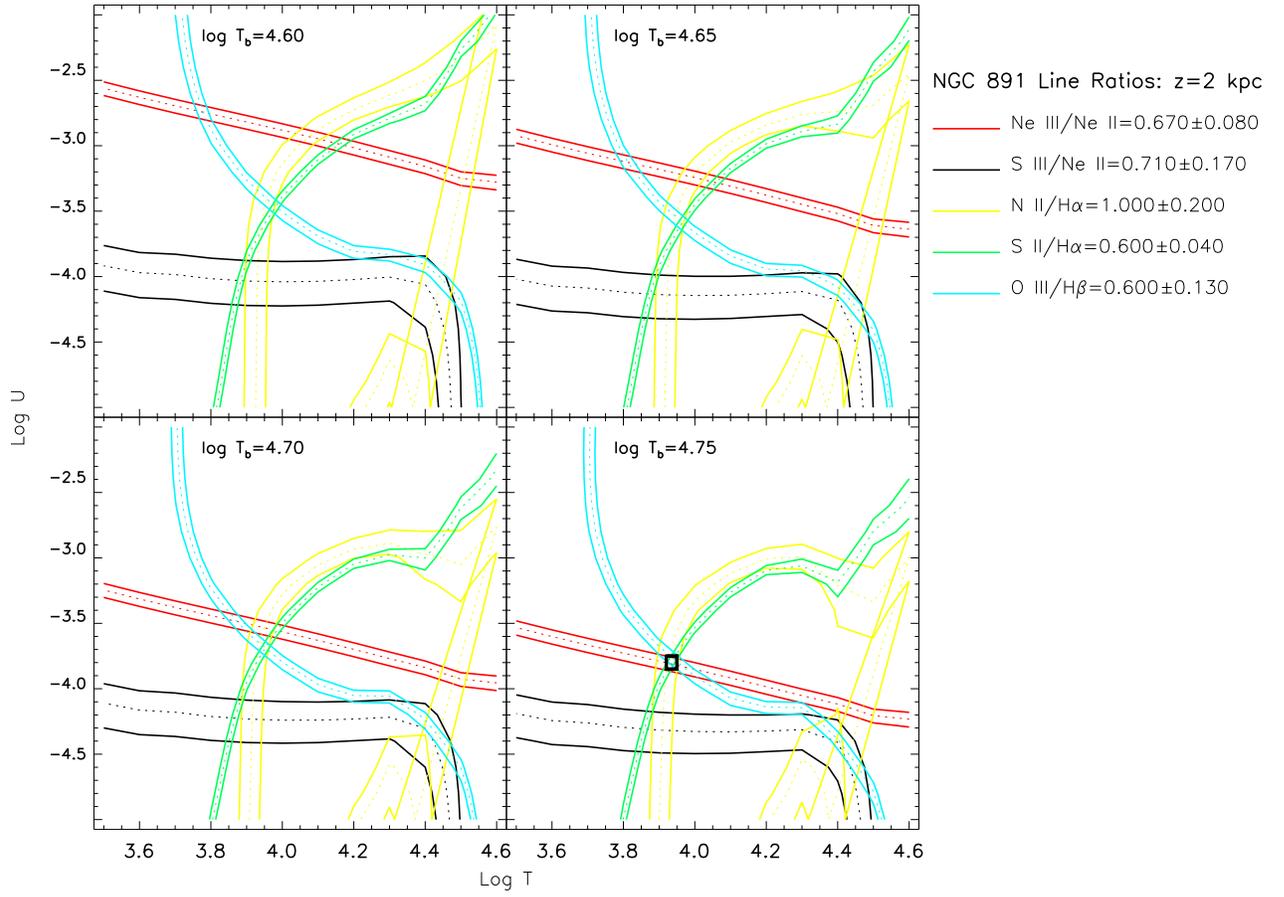}
\caption{Same as Figure 14, but for observed NGC 891 line ratios at $z=2$ kpc.}
\end{center}
\label{z2}
\end{figure}

Over the same distance, the derived gas temperature in the halo of NGC
891 increases by 20\%, from $7130\pm 820$ K to $8600 \pm 300$ K. This
modest increase is also consistent with the increase in gas
temperature that has been claimed for the Milky Way by
\citet*{1999ApJ...525L..21R}, who pointed out that this might be
evidence for a non-photoionization source of heating at low densities.
One test of this temperature gradient in edge-on systems would be a
detection of the faint [N$\,$II] 5755 \AA\ line, as the ratio of this
to the brighter [N$\,$II] line is an unambiguous temperature
diagnostic \citep{2001ApJ...548L.221R}.  So far, only 2$\sigma$ upper
limits on the temperature in the DIG halo of NGC 891 of 13,000 K and
10,000 K on the east and west sides, respectively, have been set from
the very sensitive spectrum of \citet{1997ApJ...474..129R}.  However,
another way of estimating the gas temperature, based on the
[N$\,$II]$\lambda 6583$/H$\alpha$ ratio \citep{1999ApJ...523..223H},
and applied to NGC 891 \citep{2001ApJ...551...57C}, also suggests a
rise in gas temperature consistent with the derived rise here within
the errors.

The most surprising result of our parametrized approach to
constraining the physical conditions is the required hardening of the
radiation field. Parametrized as a blackbody temperature, this is a
58\% increase, from 36,000 K to 57,000 K.  This change cannot be
explained by the radiation hardening expected as the photoionizing
flux from the base of the halo loses lower energy photons as the
radiation percolates upward (RWB). Some additional source of ionizing
radiation with a hard spectrum is needed.  However, this source of
radiation must not produce too much thermal heating of the gas, nor
may it exceed the constraints on the ionization parameter. In \S 4.3,
we explore one such possible source.

It is fortunate that this approach, with three parameters, can match four or
five line ratios simultaneously, but what about the ratios we have
neglected so far? Of the three remaining ratios, the most challenging
problem is to match the rising [O$\,$I]/H$\alpha$ at the same
time that [O$\,$III]/H$\beta$ is also rising. In general, [O$\,$I]
emission is thought to trace of the neutral phase. The neutral
oxygen fraction is "locked" to the neutral hydrogen fraction by charge
exchange, but the gas must also be hot enough to excite the optical
line emission. For that reason, it has generally been assumed that 
[O$\,$I] should trace regions of low or partial ionization, in
particular the warm boundaries of neutral clouds \citep{1998ApJ...494L..99R}.

Figures 13-15 make it clear that to match both this line ratio {\it
  and} [S$\,$III]/[Ne$\,$II], while still maintaining the
same values for [S$\,$II]/H$\alpha$ and [N$\,$II]/H$\alpha$, requires
an ionization parameter of $\log U=-4.6~(z=0\ {\rm kpc})$ or $\log
U=-4.8~(z=1\ {\rm kpc})$. If we were to assume a two-phase medium with
identical gas temperatures and irradiating spectral shape, this would
require a density ratio of 32 and 20 at $z=0$ kpc and 1 kpc,
respectively. These are plausible density contrasts for a two-phase
interstellar medium \citep{2003ApJ...587..278W}. However, this assumes
that the H$\alpha$ and [O$\,$I] emission regions are co-spatial.
\citet{1998ApJ...494L..99R} argue that in the Galactic warm ionized
medium most of the H$\alpha$ emission arises in a different physical
region than the [O$\,$I] emission. This would increase the [O
  I]/H$\alpha$ ratio in the [O$\,$I] emitting zone, requiring a
different set of parameters. (For a fixed temperature and radiation
temperature, this would yield an even lower ionization parameter.)

In a future study we will see if a two-phase or multi-phase
combination can be constructed to match {\it all} of the emission line
ratios. This will include the line ratios considered here as well as
the increasing [O$\,$II] 3727 \AA/H$\beta$ emission observed in
edge-on spirals \citep{2001ApJ...560..207O} -- including along the same
slit location in NGC 891 as considered here -- as well as the Galactic WIM
\citep{2006ApJ...650L..63M}.

\subsection{On Shocks}

It has also been suggested that shocks may play a role in increasing
the ionization level with height (e.g. \citealt{1998ApJ...501..137R},
\citealt{2001ApJ...551...57C}, \citealt{2003ApJ...592...79M}).
Although it is well established
that the power requirements to maintain these ionized layers can only
be supplied by stellar radiation, and not, for example, the mechanical
input of supernovae shocks \citep{1990ApJ...349L..17R}, this does not
preclude the possibility of shocks altering the line ratios in the
lower density, high $z$ gas where the power requirements are much
reduced. Shocks, of course, are just one example of
non-photoionization heating, where the kinetic energy of gas is
converted to thermal energy and radiation.  Other possibilities
include turbulence (kinetic energy $\rightarrow$ thermal energy + radiation),
magnetic reconnection (magnetic $\rightarrow$ thermal energy + radiation),
and cosmic-ray heating (particle kinetic energy $\rightarrow$ thermal
energy + radiation).
 
In a standard plane-parallel shock model, the optical line emission
originates in three distinct zones \citep{1996ApJS..102..161D}. First,
there is the {\it non-equilibrium cooling zone}.  As the gas behind a
shock front cools, it passes through the different ionization stages,
where the ionization level is set by the gas temperature and
by photoionization from the radiation produced by the hotter gas closer
to the shock front. However, frequently the cooling is faster than the
recombination and the ionization level is no longer in
equilibrium. Ions with the highest ionization potential are the most
likely to be produced in this zone. The second relevant region is the
{\it photoabsorption zone}. For a strong shock, the gas is compressed
by a factor of four by the shock front. As it cools, it compresses
further, where the amount of compression depends on the relative
importance of thermal and magnetic pressure. When the gas reaches
$\sim 10^{4}$K, it reaches a plateau in temperature until all of the ionizing
flux produced in the post-shock gas has been absorbed. Since this
zone is compressed, and irradiated by an extreme ultraviolet or X-ray
flux, it usually dominates the optical line emission from the shock
front, with emission line ratios distinctly different from standard
\ion{H}{2} regions. For example, [Ne$\,$III] is enhanced relative to
    [O$\,$III] by a factor of $\sim$ 6 in shocks
    \citep{1996ApJS..102..161D}. The third region is the {\it ionized
      precursor}.  Radiation from the post-shock gas streams in front
    of the shock, ionizing the low density pre-shock medium.

There are two reasons why it is difficult to compare standard shock
models to data on low density halos. First, these models are usually
calculated in the absence of an external radiation field. Since the
bulk of the emission for most of the optical/infrared lines tends to come
from the photoabsorption zone (due to the higher emission measures),
the halo radiation field must also be added self-consistently to the
radiation emitted by the post-shock gas. For example, in the low
density conditions of the halo, this photoabsorption zone may never
fully recombine.

The second difficulty is that at the low densities of the halo, the
physical thickness of the shock increases to the point that the
plane-parallel approximation becomes inappropriate. In can be shown
that the emergent hydrogen ionizing photon flux is proportional to the
particle flux through a plane-parallel shock front,
$\Phi_{H}=\phi(v_{s}) n_{H,ps} v_{s}$, where $v_{s}$ is the
shock speed, and $n_{H,ps}$ is the pre-shock density. The dimensionless
efficiency factor, $\phi(v_{s})$, depends principally on the shock
speed and modestly on the amount of compression allowed by the
magnetic field in the post-shock gas. This factor ranges from 3 to 12
for shock speeds between 150 to 300 ${\rm km~s^{-1}}$
\citep{1994PhDT........52B}. This allows us to estimate the length of
the precursor zone and the photoabsorption zone by assuming the
ionizing photon flux incident on each zone is balanced by the
recombination in a gas column, $l=\Phi_{H}/(n_{H}^{2} \alpha)=(\phi
v_{s})/(n_{H}\alpha)=(1200 {\rm pc})~ (\phi\,n_{-2})^{-1}\,v_{100}\,
T_{4}^{0.7}$, where $T_{4}$ is the gas temperature in units of
$10^{4}$ K, $v_{100}$ is the shock velocity in units of $100~{\rm
  km~s^{-1}}$, $\alpha=(2.6 \times 10^{-13}~{\rm
  cm^{3}~s^{-1}})T_{4}^{-0.7}$ is the hydrogen recombination rate, and
$n_{-2}$ is the hydrogen density in the irradiated gas in units of
$10^{-2}~{\rm cm^{-3}}$. We see that for reasonable halo parameters,
the thickness of the precursor zone can exceed the thickness of the
entire ionized layer. For this reason, many shock model predictions
exclude the contribution of the precursor. The thickness of the
photoabsorption zone will be reduced by the compression factor
experienced by the post-shock gas, typically by a factor of 4 to 400.
The compression depends principally on the magnetic field in the
post-shock gas.

Because of the certain presence of the external radiation field, and
the fact that the bulk of the optical line emission in shocks is
reprocessed flux from the post-shock gas, our previous approach of
parametrizing the emission as a function of temperature, ionization
parameter, and spectral hardness is still appropriate. However, the
spectral shape will be much different than the blackbody flux that we
have used. In principle, we could repeat the analysis in the previous
subsection, replacing the blackbody spectrum with the spectrum
emergent from shock-heated gas {\it combined} with the background
stellar flux, and we will consider this in future work. This would
require the incorporation of (at least) two additional free
parameters, the shock velocity/post-shock temperature and the flux
level of ionizing photons.

\include{tab1}

\subsection{Illustrative Physical Models for NGC 891}

In RWB we used the Monte Carlo photo-ionization code described in WM
to attempt to understand whether the rise of the neon ratio with $z$
in NGC 891 found in our {\it Spitzer} Cycle 2 data could be explained
by photo-ionization due to massive stars in a thin disk.  We also
considered the runs of the optical line ratios shown in Figure 11.
The details of the models are presented in RWB, therefore we emphasize
here only a few of the most important features.  Stars of temperature
40 kK, 45 kK and 50 kK were considered, with a total luminosity of
10$^{51}$ ionizing photons per second, although one model considered a
luminosity four times higher.  While our standard models place the
stars in a diffuse gas density structure (based on observations of NGC
891), one model also considers the effects of spectral hardening by
denser gas around the stars - this involves passing the radiation
field through a spherical gas distribution whose size is set to allow
15\% of the ionizing photons to escape.

Our main result was that no model could reproduce the rise of
[Ne$\,$III]/[Ne$\,$II] with z, with values of this ratio being
generally overpredicted.  The models did a better job in reproducing
the observed runs of [S$\,$II]/H$\alpha$ and [N$\,$II]/H$\alpha$,
but could not produce a rising [O$\,$III]/H$\beta$ as observed,
predicted too low values of [O$\,$I]/H$\alpha$, and could not match
the values of He I/H$\alpha$.

The additional data on NGC 891 presented in this paper and the above
exploration of parameter space in photo-ionization scenarios with
CLOUDY show that the situation will not change -- a thin disk of
hot stars will not suffice.  Instead they suggest that we vary the
distribution of ionizing sources, such that gas at higher $z$ is
ionized primarily by sources that have, in the mean, harder spectra.
Here we focus on such a source which is well suited to modeling by the
Monte Carlo code by adding a secondary, hot stellar ionizing source
with a vertically extended distribution.

Our goal here is simply to present some illustrative models showing
parameters for such a component (temperature, scale height, fraction
of ionizing photons contributed) that, when added to our previous
models, can approximately match the neon ratio behavior, rather than
to perform an exhaustive search of parameter space.  The first reason
for this limited approach is that there are many possible ways in
which the parameters suggested by our CLOUDY models could be
incorporated into a physical model, not all of which may have a
well-motivated physical explanation.  Second, while we can find
motivation for including the second source in several candidate types
of stars based on Milky Way studies, information is scarce concerning
their potential effectiveness as contributors to the ionization, and
virtually non-existent regarding their distribution in NGC 891.
Hence, our problem is not well constrained.

Hot sources in the Milky Way that may have a larger vertical
extent than the main massive star layer include sdO and sdB stars,
post-AGB stars, young white dwarfs, and runaway O stars.  We return to
the viability of each of these candidates after we present the
modeling results.  However, we do note here two related contexts in
which such sources have been considered: \citet{1994A&A...292...13B},
\citet*{2000AJ....120.1265T}, and \citet{2008MNRAS.391L..29S} have
explored possible importance of hot post-AGB stars and planetary
nebula nuclei in explaining high [O$\,$III]/H$\beta$ and
[N$\,$II]/H$\alpha$ ratios in Seyfert and LINER spectra, while
\citet{2010arXiv1004.1647A} and \citet{2010MNRAS.402.2187S} have
considered the importance of post-AGB stars for ionization in
elliptical galaxies.


Instead of the point-source geometry for the ionizing sources
considered in RWB, here we use plane-parallel distributions.  The
massive stars occupy an infinitely thin layer, while an exponential
describes the vertical distribution of the secondary component.  We
simulate a disk area of 4 kpc by 4 kpc.  For the massive stars, we
only consider a temperature of 30 kK, and do not consider leakage.
For the second component, we focus on hot, evolved, low mass stars and
use stellar spectra from the T${\ddot {\rm u}}$bingen NLTE model
atmospheres \citep[e.g.][]{2009ASPC..411..388R}.  We consider 50 kK
models with elements from H to Ni, as well as ones with no H
(He+C+N+O) to represent He-rich stars such as the sdOs.  This choice
should give us at least some idea of whether our results are sensitive
to composition, which can vary greatly among subdwarf stars, for
example \citep{2009ARA&A..47..211H}.  The total ionizing luminosity in
our models is $5.5 \times 10^{50}$ photons s$^{-1}$.

Figures 16 and 17 show resulting line ratios for two models where
stars with the H-Ni spectra with a scale height of 1 kpc contribute
5\% and 7.5\% (``HNi5'' and ``HNi7.5'' models) of the total ionizing
luminosity, respectively.  The extended component has only been
included on the positive-$z$ side, although it will also slightly
ionize the opposite side as the emission is isotropic.  First,
we note that almost all line ratios show a sharp spike or dip in the
midplane which is not seen in the data.  This is a consequence of the
idealized nature of the models, where the midplane line of sight
passes through the center of the ionized structure.  Real lines of
sight in the midplane presumably sample a range of such impact
parameters, and should therefore reflect some mix of DIG and
HII regions.  Hence, we do not expect to match ratios correctly within
the star forming layer.

\clearpage
\begin{figure}
\epsscale{.80}
\includegraphics[scale=.8,viewport=0 0 574 668]{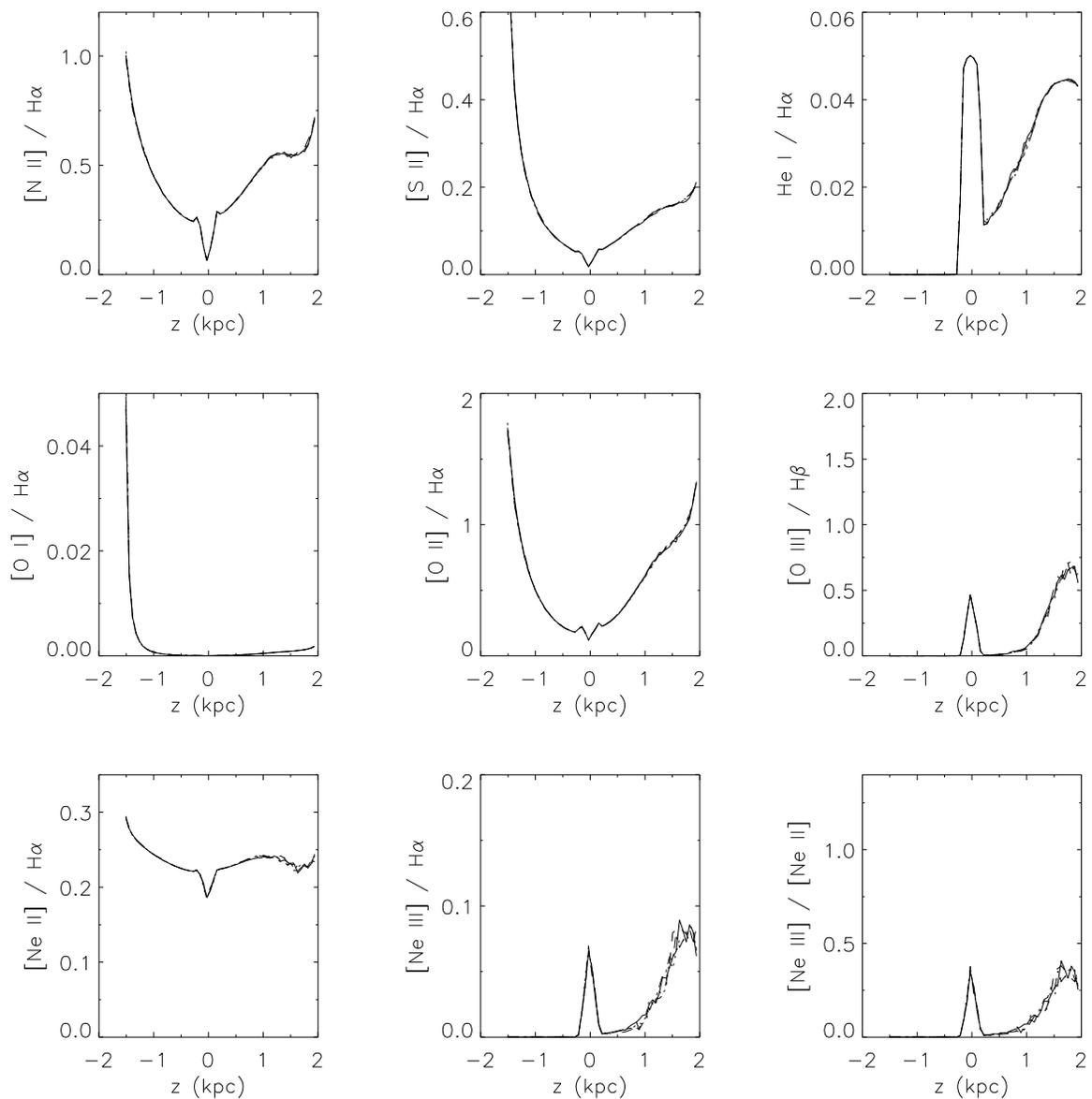}
\caption{Vertical cuts showing the variation of line ratios with $z$
  for the ``HNi5'' model (for doublets, the same line is plotted as in
  Figure 11).  The vertically extended ionizing source is located on
  the positive $z$ side only.  The different curves in each panel
  represent cuts through the model centered at various in-plane
  coordinates, and reflect the noise in the simulation.}
\label{fig16}
\end{figure}
\clearpage

\clearpage
\begin{figure}
\epsscale{.80}
\includegraphics[scale=.8,viewport=0 0 574 668]{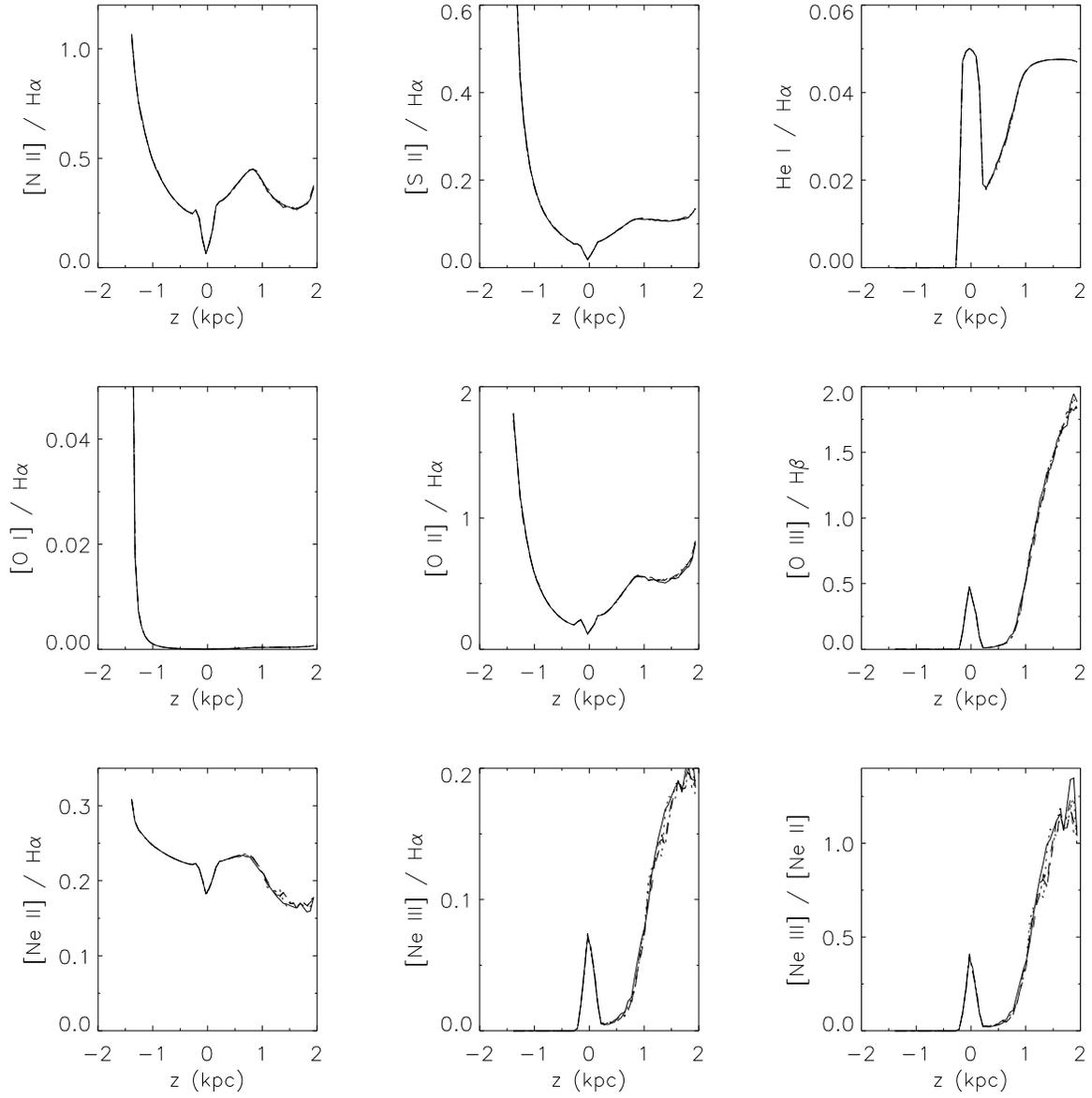}
\caption{Same as Figure 16, but for the ``HNi7.5'' model.}
\label{fig17}
\end{figure}
\clearpage

At $z=1$ and 2 kpc, these models bracket the observed neon line ratios
seen in NGC 891.  Varying the contribution from the extended stellar
component has little effect on the behavior of [Ne$\,$II]/H$\alpha$
with height, but greatly affects the behavior of [Ne$\,$III]/H$\alpha$
(we do not report these ratios for the data mainly because extinction
in the infrared and optical lines should vastly differ).  Models with
scale heights of 0.5 kpc and 2 kpc (not shown) produce, respectively,
no rise, and much too steep a rise in the neon ratio off the plane.
From this we can conclude that, for this representative temperature of
the extended component, the rate of rise of the neon ratio between
$z=1$ and 2 kpc is governed mainly by the scale height of these stars,
rather than their relative contribution to the ionizing flux.  Hence,
the only way to reproduce a rising neon ratio with such models is to
include a vertically extended ionizing component, as we have done.

Optical line ratios are more difficult to interpret, as already stated
-- nevertheless, we summarize here how these models fare in
reproducing these ratios.  Comparing Figures 11, 16, and 17, the HNi5
model produces the rising trend in [O$\,$III]/H$\beta$ in the halo
with values close to those observed, and comes reasonably close to
matching the observed range of values of He I/H$\alpha$ of about
$0.02-0.045$.  In these aspects, the model is much more successful
than the ones presented in RWB.  However, [S$\,$II]/H$\alpha$ and
[N$\,$II]/H$\alpha$, although rising in the model, are now too low
compared to the data.  The models of RWB were more successful in this
regard.  [O$\,$I]/H$\alpha$ is also badly underpredicted, although
this was also true for the RWB models.  A negative abundance gradient
with height will not help the situation: for instance, from the
results of WM, [S$\,$II]/H$\alpha$ and [N$\,$II]/H$\alpha$ would be
further lowered.


The HNi7.5 model, while also producing rising [S$\,$II]/H$\alpha$,
[N$\,$II]/H$\alpha$ and [O$\,$III]/H$\beta$ ratios, underpredicts
the first two even more than the HNi5 model, and overpredicts the
third.  The behavior of He I/H$\alpha$ is similar to that for the HNi5
model.

The He-rich star model (not shown) produces a very similar behavior of the
neon ratio to the HNi5 model when such stars contribute 30\% of the
ionizing flux.  Their contribution is necessarily relatively large
because the high helium content of the stars' atmospheres suppresses
the neon-ionizing flux significantly.  However, this model yields very
low [N$\,$II]/H$\alpha$ ($0.1-0.4$) and [S$\,$II]/H$\alpha$ ($<0.1$)
ratios that decline with $z$.  [O$\,$III]/H$\beta$ rises with $z$ but
is overpredicted, [O$\,$I]/H$\alpha$ is badly underpredicted, while
the behavior of He I/H$\alpha$ is similar to that for the HNi5 and HNi7.5
models.  We therefore do not consider such models further.


If we focus only on the success of the HNi5 and HNi7.5 models in bracketing
the observed behavior of the neon ratio with height,
and ignore the problems introduced in matching other line ratios, we
can then ask which type of aforementioned hot star could be
responsible.  The distribution of such stars is unknown in NGC 891, 
and therefore we can only try to constrain the possibilities based on
what is known in the Milky Way.  The only information available on the
distribution of stars in NGC 891 relative to the Milky Way is very
general -- thin and thick disk scale heights have been measured for
both galaxies -- and we end by summarizing these measurements.

Of the various types of hot stars mentioned above, sdO stars may be
the most likely in terms of temperature and z-distribution, but have
the disadvantage that they tend to be He-rich.  These stars have
T=40--80 kK, although in a sample of 21 He-rich sdOs, most were found
to have T=45--55 kK \citep{1994ApJ...433..819T}, while in another
sample of 33 such stars, most were in the range 38--48 kK
\citep{2007A&A...462..269S}.  Temperatures for a sample of 13 sdOs
with subsolar helium abundances are in the range 35--80 kK
\citep{2007A&A...462..269S}.  Heights above the plane for the sample
of \citet{1994ApJ...433..819T} are estimated to be 0.4--1.5 kpc, and
Thejll et al. speculate that they may form a thick disk or halo
population.  More recently, \citet{2008ASPC..392..139N} finds from
kinematics of 31 He-sdOs that 41\% are in the thin disk, 39\% in the
thick disk, and 20\% in the halo.  Their space density is very
uncertain.  \citet{1986ApJS...61..569D} found a value in the Solar
neighborhood of about $7 \times 10^{-7}$ pc$^{-3}$, although with a
low scale height of $100-200$ pc.  If the scale height is indeed 1
kpc, the midplane space density would be reduced.  The midplane space
density of such stars in our HNi5 model can be derived if we
assume that our extended component consists of stars of a typical
sdO luminosity of 150 L$\sun$, based on values from
\citet{1994ApJ...433..819T}.  We find a value of $6 \times 10^{-7}$ pc$^{-3}$.
Hence it seems, although information is very scarce, that such an extended
component cannot be ruled out on grounds of space density, temperature
or scale height.

\citet{1997A&A...327..577D} find that the sdB stars also have a large
scale height of about 1 kpc, but these stars are certainly too cool to
be responsible for the neon line behavior.  The Milky Way post-AGB
stars in the catalog of \citet{2007A&A...469..799S} are also too cool,
with temperatures not exceeding 25 kK.  Much hotter post-AGB stars
include the PG1159 stars \citep*{1986ApJS...61..305G}, which have
temperatures of order 100 kK.  Only 40 are known
\citep{2008ASPC..391..109W}.

A sample of white dwarfs in the Sloan Digital Sky Survey are found to
have a scale height of about 340 pc, consistent with several earlier
determinations \citep{2006AJ....131..571H}.  These authors also find a
possible trend of increasing scale height with decreasing luminosity
(and therefore temperature), but the statistical significance is
uncertain.  Such a trend would naturally be expected because of
heating of the disk over time.  Nevertheless, it seems that luminous,
hot white dwarfs should have a scale height of much less than 1 kpc.
If the PG1159 stars are evolving into hot white dwarfs, the same
conclusion should apply to them.

Runaway O stars account for $10-30\%$ of all O stars
\citep{1987ApJS...64..545G} and form a distribution more vertically
extended than the thin disk.  However, most have $V_{lsr}$'s of less
than 100 km s$^{-1}$, so it is difficult to see how they could form a
layer with a 1 kpc scale height given their lifetimes.  In fact, such
stars in the sample of \citet{1998A&A...331..949M} are mostly within
200 pc of the midplane.  Some B stars are found at greater heights
\citep*[][and references therein]{2004RMxAC..21..121A}, but again are
too cool to be candidates here.

Of course, candidate sources with appropriate temperatures may have a
different scale height and volume density in NGC 891.  While
individual candidate stellar types are difficult to study, the overall
stellar disk structure of NGC 891 has been modeled several times,
beginning with \citet{1981A&A....95..116V}, who were the first to
report a thick disk.  From $R$-band photometry,
\citet{1997AJ....113.2061M} derived scale heights for the thin and
thick disks of $400-650$ pc and $1.5-2.5$ kpc, respectively, with the
thick disk contributing $5-10\%$ of the surface brightness in the
midplane.  \citet{1998A&A...331..894X}, modeling starlight and
extinction from five-band photometry, found a thin disk scale height
of 400 pc.  More recently, \citet*{2009MNRAS.395..126I} found a thick
disk scale height of $1.44 \pm 0.03$ kpc from star counts in HST ACS
images.  Thin and thick disk scale heights are therefore larger than in
the Milky Way, where \citet{2008ApJ...673..864J} find values of 0.3
kpc and 0.9 kpc from the Sloan Digital Sky Survey.  Hence, there may
well be an extended distribution of hot, evolved stars in NGC 891, but
one cannot currently say more than this.  

We conclude from this initial exploration that such a component cannot
be ruled out as an explanation for the neon ratio behavior given what
is known about the stellar distribution in NGC 891 and the required
space density of this component in our models.  It also may help
explain the [O$\,$III]/H$\beta$ and He I/H$\alpha$ behavior in this
galaxy.  However its inclusion does introduce problems with other
optical line ratios.  As previously mentioned, a more complete
exploration of parameter space in such models, including a
consideration of non-ionizing heating, may result in a better match to
the data for NGC 891, but is beyond the scope of this paper.

Finally, we note that any contribution of a hidden AGN or nuclear
super star cluster with a hard spectrum is very unlikely to be a
factor in NGC 891.  First, the structure of the H$\alpha$ halo
emission correlates with star formation in the disk, both in the
contrast on either side of the minor axis, and on the scale of
individual filaments \citep{1998PASA...15..106R}.  Second, in a slit
running parallel to the disk at a height of 700 pc in the DIG halo,
[N$\,$II]/H$\alpha$, [S$\,$II]/H$\alpha$, and [O$\,$I]/H$\alpha$ all
anti-correlate with H$\alpha$ halo intensity, dropping at the location
of filaments \citep{1998ApJ...501..137R}.  These results imply a
strong connection between line ratios and disk star formation, with no
indication of a contribution from a nuclear source.

\section{Discussion of PAHs}

Our main results for PAH emission scale heights are as follows.
First, the scale heights (depending on extinction) are many hundreds
of pc.  This result adds to the growing evidence for vertically
extended PAH distributions in galaxies (e.g.,
\citealt{2006ApJ...642L.127E}; \citealt{2008ApJ...676..304B};
\citealt{2009arXiv0904.4928H}).  Specifically for these two galaxies,
in the IRAC 8 $\mu$m image of NGC 891, \citet{2009MNRAS.395...97W}
measure a scale height of $248 \pm 50$ pc (with no extinction
correction), while the IRAC 8 $\mu$m image of NGC 5775 presented by
\citet{2009arXiv0904.4928H} clearly shows a very extended layer with
much vertical filamentary structure -- although this wavelength traces
more ionized PAHs.  Second, the PAH scale heights are greater in NGC
5775 than in NGC 891 -- the former also featuring more active star
formation and more extended warm ionized and neutral gas halos
(\citealt{1994ApJ...429..618I}; \citealt{2000ApJ...536..645C}; 
\citealt{2007AJ....134.1019O}; \citealt{1990ApJ...352L...1R}).  Our PAH
scale heights are comparable to or somewhat lower than the HI scale
heights.  Third, within NGC 891 and NGC 5775, scale heights are similar
among the various PAH features.

There are two obvious caveats to consider when interpreting these PAH
scale heights.  The first is that we have halo
measurements at only one distance along the major axis in each galaxy.
The second is that the scale heights should reflect the intensity of
the exciting radiation as well as the PAH density distribution.

No PAH feature stands out as having a significantly greater or smaller
scale height in either galaxy halo.  Thus, there is little evidence
for significant modification between disk and halo of the relatively
large, neutral PAH population responsible for the discrete features in
this wavelength range.  However, for NGC 891, the 8 $\mu$m (dominated
by the PAH feature at 7.7 $\mu$m) scale height of $248 \pm 50$ pc
measured by \citet{2009MNRAS.395...97W} (uncorrected for extinction)
is significantly lower than our average of $475 \pm 30$ pc for the PAH
features.  In order for this difference to be an extinction effect,
the extinction at 8 $\mu$m would need to be about 1 mag greater than
for our PAH features.  Although 9.7 $\mu$m silicate absorption may
have a significant effect in the 8 $\mu$m band, it may also affect the
11.2 $\mu$m band (perhaps even more so than the 7.7 $\mu$m feature
which dominates the 8 $\mu$m images; \citealt{2008ApJ...679..310G}),
while the 17.4 $\mu$m feature could suffer from significant, albeit
less, extinction due to the 18 $\mu$m silicate feature.  We therefore
consider extinction an unlikely explanation.  Furthermore, the PAH
emission spectrum should be relatively independent of ISRF strength
for the weak radiation fields expected in integrated disk and halo
lines of sight (\citealt{2008ApJ...679..310G}; \citealt
{2007ApJ...657..810D}).  Therefore the difference in scale heights
most likely represents a real change in the PAH population between
disk and halo.  The result seems consistent with the drop of the
7.7 $\mu$m/11.2 $\mu$m intensity ratio with $z$ in the M82 outflow
\citep{2008ApJ...679..310G}.  According to the models of
\citet{2007ApJ...657..810D}, the 7.7 $\mu$m feature is due to smaller
and more ionized PAHs than those responsible for the $10-20$ $\mu$m
features, although \citet{2008ApJ...679..310G} claim that the 7.7 
$\mu$m/11.2 $\mu$m ratio is sensitive only to ionization and not size
variations.  Our results would then indicate a significant drop in PAH
ionization, and possibly a transition to larger PAHs, with height in
the halo of NGC 891.

These results suggest a picture in which PAHs are transported upward
from the disk (where they are thought to form in C-rich evolved AGB
star envelopes) coupled with warm gas as part of a disk-halo flow.
However, as discussed in \S 1, other mechanisms could be at work.
Regarding radiative acceleration of dust, \citet{1998MNRAS.300.1006D}
find that the process is less efficient for smaller grains.
Alternatively, PAHs could form {\it in situ}.  One possible mechanism is the
shattering of larger grains in shocks, which could be effective for
shock speeds as low as 50 km s$^{-1}$ (\citealt{1996ApJ...469..740J}).

Shocks may also destroy PAHs, with larger ones being able to withstand
faster shocks.  \citet*{2010A&A...510A..36M} find partial destruction
of PAHs for shocks of $75-100$ km s$^{-1}$, while ones with 50 and 200
C atoms do not survive shocks faster than 100 km s$^{-1}$ and 125 km
s$^{-1}$, respectively.  PAHS in the $10-20$ $\mu$m range probably
contain a few hundred to a few thousand C atoms
\citep{2007ApJ...657..810D}, so if shock speeds are less than about
125 km s$^{-1}$
they should survive.  On
the other hand, the lower scale height of the 8 $\mu$m emission
relative to the $10-20$ $\mu$m features could be due to shock
modification.  If the former arise mostly from PAHs of $\sim 100$ C
atoms \citep{2007ApJ...657..810D} there may be partial destruction
for shock speeds near 100 km s$^{-1}$, if one interpolates the models of
\citet{2010A&A...510A..36M} for the 50 and 200 C atom cases.

We have also found that most $10-20$ $\mu$m PAH EWs are higher at
$z=1-2$ kpc than in the midplane, using two different methods to
estimate the continuum.  \citet{2008ApJ...676..304B} find a similar
trend between disk and halo for PAH EWs in this wavelength range in
M82.  One might conclude that, together with the increase in
[Ne$\,$III]/[Ne$\,$II] seen in NGC 891 and NGC 5775, the result
reflects previous findings of an increase of EWs with radiation field
hardness (e.g., \citealt{2006ApJ...639..157W} for Blue Compact Dwarf
Galaxies; \citealt{2008ApJ...678..804E} for starbursts).  However, our
values are in the regime [Ne$\,$III]/[Ne$\,$II] $<1$, where such a
correlation does not exist in star forming regions
(\citealt{2006ApJ...653.1129B}; \citealt{2008ApJ...682..336G}).  PAH
features also weaken in galaxies with low metallicity (e.g.,
\citealt{2006A&A...446..877M}; \citealt {2006ApJ...639..157W}).  There
is no direct information on how metallicity varies with height in our
three halos.  Some lines of evidence, for instance the generally low
metallicities of Milky Way High Velocity Clouds
(\citealt{2004Ap&SS.289..381W}), suggest halos may have lower
metallicities than disks, possibly as a result of mixing of disk-halo
cycled and primordial gas.  However, these clouds are found at heights
much greater than $1-2$ kpc (\citealt{2007ApJ...670L.113W}).
Regardless, a drop in metallicity with height would be expected to
produce, if anything, a drop in EWs.

Hence, the explanation of this result is not clear.  As mentioned
before, it should be kept in mind that there is evidence
(\citealt{2007ApJ...663..866D}) from SINGS galaxies that, just like
the discrete features, the continuum in the 16 $\mu$m range is mainly
due to single-photon heating of PAHs.  In fact, feature and continuum
radiation in this range is dominated by molecules of size $5-20\AA$
molecules for weak radiation fields in the model of
\citet{2007ApJ...657..810D}.  If the continuum throughout the $10-20$ 
$\mu$m range is also dominated by single-photon PAH heating, then our
result suggests a perhaps rather subtle change in the PAH population
between disk and halo.  It is not clear which, if any, of the physical
mechanisms described above may be responsible.



\section{Conclusions}

We have presented $10-20$ $\mu$m spectroscopy of the disks and halos of
the edge-on galaxies NGC 891, NGC 5775 and NGC 3044.  Regarding the
gas-phase lines, our main result is that [Ne$\,$III]/[Ne$\,$II] is
higher in the halos of NGC 891 and NGC 5775 than in the disks.
Scatter in NGC 3044 prevents any trend from being seen.

These results exacerbate the problem of explaining DIG line ratio
behavior with simple photo-ionization models featuring a thin disk of
massive stars.  Focusing on NGC 891, where we have the most
observational constraints, we have explored parameter space in CLOUDY
photo-ionization models to try to determine, as a function of height,
what combination of radiation temperature (assuming blackbodies for
simplicity), ionization parameter and gas temperature may reproduce
the observed IR and optical line ratios.  We find that a dramatic rise
in radiation temperature with $z$ is required, as well as a more
modest rise in gas temperature and a fall in ionization parameter.
Not all line ratios can be reproduced without considering more complex
models, however.  We then considered representative physical
photo-ionization models incorporating a thin disk of massive stars in
combination with a thick disk of hot (50 kK), presumably evolved
stars.  We found that the neon ratio behavior in particular could be
approximately reproduced for such a thick disk with scale height of
order 1 kpc and a contribution of $5-7.5\%$ of the ionizing radiation.
Whether such a component exists in NGC 891 (or even the Milky Way)
remains very unclear, but it cannot yet be ruled out
based on space density, temperature and scale height arguments.  In
our models, such a component is also reasonably successful in
reproducing the rising [O$\,$III]/H$\beta$ and approximate He
I/H$\alpha$ values in this galaxy.  However its inclusion does
introduce problems with other line ratios.  A secondary stellar source
is not the only way of increasing radiation hardness with height, and
our exploration of parameter space with CLOUDY suggests a feasible way
in which other such sources, including hard EUV/X-ray radiation, may
be included in future work.

PAHs in NGC 891 and NGC 5775 form a vertically extended distribution,
with emission scale heights larger in NGC 5775, as is the case for the
HI gas.  The PAH emission scale heights are comparable to, or somewhat
less than, the HI values.  These results suggest that PAHs participate
in the active disk-halo flows in these two galaxies.  In NGC 891, the
scale heights are larger than the 8 $\mu m$ scale height, suggesting a
transition to more neutral, and possibly larger, grains with $z$.
Most PAH equivalent widths are higher in the halos than in the disks,
again indicating some slight modification in the PAH population in the
halo environment.  Shocks and radiative acceleration are processes that may
affect PAH populations in halos.

The 17 $\mu$m H$_2$ line is detected in all pointings, indicating a
warm molecular gas component with a large vertical extent, reaching as
high as $z=2$ kpc.  The greater scale height of this emission in NGC
5775 {\it vs.} NGC 891 again suggests a connection with disk-halo flows.
Column densities and temperatures cannot be derived without
measurement of additional transitions, however.  Future observations
will no doubt reveal the fraction of halo gas mass in this phase, and
its importance for understanding the ISM of halos.

Finally, we wish to stress here the importance of the neon ratio in
understanding the energetics of the diffuse ISM, and the great
advantage it holds over optical diagnostics.  Future instruments
capable of sensitive, high spatial and spectral resolution
observations of these and other mid-IR lines should be a strong
priority as these lines reveal a wealth of information.


\acknowledgments

We thank J. Mathis for inspiring us to reexamine the issue of
hot, evolved stars as potential contributors to DIG ionization.  
We also thank B. Draine for helpful comments regarding PAHs, and
J. Irwin for providing CO 2--1 data for NGC 3044.

RAB would like to acknowledge the support of
Spitzer Space Telescope Guest Investigator award, NASA/JPL Contract
1324695, and the NASA Astrophysical Theory grant NNX10AI70G to the
University of Wisconsin-Whitewater.

This work is based (in part) on observations made with the Spitzer
Space Telescope, which is operated by the Jet Propulsion Laboratory,
California Institute of Technology under a contract with NASA. Support
for this work was provided by NASA through an award issued by
JPL/Caltech.



Facilities: \facility{Spitzer Space Telescope}.




\bibliography{ms}

\clearpage




\begin{deluxetable}{ll}
\tabletypesize{\scriptsize}
\tablecaption{Emission Lines and Ionization Potentials\label{tbl-1}}
\tablewidth{0pt}
\tablehead{
\colhead{Line}   & 
\colhead{Ionization Potential (eV)}
}
\startdata

H$\alpha$ 6563$\AA$ & 13.6 \\
$[$S$\,$II$]$ 6716$\AA$ & 10.4 \\
$[$N$\,$II$]$ 6583$\AA$ & 14.5 \\
$[$O$\,$I$]$ 6300$\AA$ & 13.6 \\
$[$O$\,$III$]$ 5007$\AA$ & 35.1 \\
He$\,$I 5876$\AA$ & 24.6 \\
$[$Ne$\,$II$]$ 12.81 $\mu$m & 21.6 \\
$[$Ne$\,$III$]$ 15.56 $\mu$m & 41.0 \\
$[$S$\,$III$]$ 18.71 $\mu$m & 19.1 \\
$[$S$\,$IV$]$ 10.51 $\mu$m & 34.8 \\
\enddata


\end{deluxetable}
\clearpage

\begin{deluxetable}{lllll}
\tabletypesize{\scriptsize}
\tablecaption{IRS Observations\label{tbl-2}}
\tablewidth{0pt}
\tablehead{
\colhead{Date}   & 
\colhead{Pointing\tablenotemark{a}} & 
\colhead{R.A. (J2000.0)} & 
\colhead{Decl. (J2000.0)} & 
\colhead{Integ. time per nod (sec)}
}
\startdata

 2008 Mar 26 &   NGC 5775 Disk 1   & 14$^{\rm h}$ 53$^{\rm m}$ 58.4$^{\rm s}$ & 3$^{\circ}$ $32\arcmin$ 24$\arcsec$ & 975 \\
 2008 Mar 26 &   NGC 5775 Disk 2   & 14$^{\rm h}$ 53$^{\rm m}$ 59.0$^{\rm s}$ & 3$^{\circ}$ $32\arcmin$ 12$\arcsec$ & 975 \\
 2008 Mar 26 &   NGC 5775 Disk 3   & 14$^{\rm h}$ 53$^{\rm m}$ 59.6$^{\rm s}$ & 3$^{\circ}$ $32\arcmin$ 00$\arcsec$ & 975 \\
 2008 Mar 26 &   NGC 5775 Halo     & 14$^{\rm h}$ 53$^{\rm m}$ 57.5$^{\rm s}$ & 3$^{\circ}$ $32\arcmin$ 14$\arcsec$ & 3854 \\
 2008 Mar 26 &   NGC 5775 Sky      & 14$^{\rm h}$ 54$^{\rm m}$ 11.7$^{\rm s}$ & 3$^{\circ}$ $35\arcmin$ 58$\arcsec$ & 3854 \\
 2008 Sep 9 &   NGC 891 Halo east  & 2$^{\rm h}$ 22$^{\rm m}$ 40.5$^{\rm s}$ & 42$^{\circ}$ $22\arcmin$ 13$\arcsec$ &  21194 \\
 2008 Sep 9 &   NGC 891 Halo west  & 2$^{\rm h}$ 22$^{\rm m}$ 33.2$^{\rm s}$ & 42$^{\circ}$ $22\arcmin$ 44$\arcsec$ &  21194 \\
 2008 Sep 9 &   NGC 891 Sky        & 2$^{\rm h}$ 23$^{\rm m}$ 32.7$^{\rm s}$ & 42$^{\circ}$ $22\arcmin$ 21$\arcsec$ &  21194 \\
 2009 Jan 7 &  NGC 3044 Halo SE   & 9$^{\rm h}$ 53$^{\rm m}$ 39.1$^{\rm s}$ & 1$^{\circ}$ $34\arcmin$ 42$\arcsec$  & 10597 \\
 2009 Jan 7 &  NGC 3044 Disk 1    & 9$^{\rm h}$ 53$^{\rm m}$ 38.6$^{\rm s}$ & 1$^{\circ}$ $35\arcmin$ 01$\arcsec$  & 975 \\
 2009 Jan 8 &  NGC 3044 Disk 2    & 9$^{\rm h}$ 53$^{\rm m}$ 39.5$^{\rm s}$ & 1$^{\circ}$ $34\arcmin$ 55$\arcsec$  & 975 \\
 2009 Jan 8 &  NGC 3044 Disk 3    & 9$^{\rm h}$ 53$^{\rm m}$ 40.4$^{\rm s}$ & 1$^{\circ}$ $34\arcmin$ 49$\arcsec$  & 975 \\
 2009 Jan 8 &  NGC 3044 Halo NW   & 9$^{\rm h}$ 53$^{\rm m}$ 39.9$^{\rm s}$ & 1$^{\circ}$ $35\arcmin$ 08$\arcsec$  & 10597 \\
 2009 Jan 8 &  NGC 3044 Sky       & 9$^{\rm h}$ 53$^{\rm m}$ 39.5$^{\rm s}$ & 1$^{\circ}$ $36\arcmin$ 30$\arcsec$  & 10597 \\
\enddata


\tablenotetext{a}{Pointing center of the field of view (each nod
pointing is symmetrically offset from this position)}

\end{deluxetable}
\clearpage

\begin{deluxetable}{rcccc}
\tabletypesize{\scriptsize}
\tablecaption{NGC 891 Infrared Line Intensities and PAH Equivalent Widths\label{tbl-3}}
\tablewidth{0pt}

\tablehead{
\colhead{} & 
\colhead{East Halo} & \colhead{East Halo} & \colhead{West Halo} & \colhead{West Halo} \\
\colhead{Line} & \colhead{Intensity\tablenotemark{a}} & \colhead{EW} & \colhead{Intensity} & \colhead{EW} \\
\colhead{} &
\colhead{} &
\colhead{$\mu$m} &
\colhead{} &
\colhead{$\mu$m} \\
}
\startdata

\phd [S$\,$IV]   10.51 $\mu$m&$<$0.17\tablenotemark{b}&~&$<$0.16&~\\
\phd [Ne$\,$II]  12.81 $\mu$m&2.4$\pm$0.2&~&3.0$\pm$0.2&~\\
\phd [Ne$\,$III] 15.56 $\mu$m&1.6$\pm$0.1&~&1.4$\pm$0.1&~\\
\phd [S$\,$III]  18.71 $\mu$m&1.7$\pm$0.1&~&1.0$\pm$0.1&~\\
\phd H$_2$ {\it S}(1) $J=3-1$ 17.03 $\mu$m&1.4$\pm$0.1&~&1.7$\pm$0.1&~\\
\phd PAH 11.2 $\mu$m&18.2$\pm$1.3&1.14$\pm$0.04&19.2$\pm$1.4&1.36$\pm$0.05\\
\phd PAH 12.0 $\mu$m&1.6$\pm$0.3&0.10$\pm$0.02&2.2$\pm$0.3&0.15$\pm$0.02\\
\phd PAH 12.7 $\mu$m&7.1$\pm$0.5&0.52$\pm$0.02&7.8$\pm$0.6&0.64$\pm$0.02\\
\phd PAH 16.5 $\mu$m&0.8$\pm$0.1&0.11$\pm$0.02&1.2$\pm$0.1&0.23$\pm$0.02\\
\phd PAH 17.4 $\mu$m&0.4$\pm$0.1&0.07$\pm$0.02&0.7$\pm$0.1&0.13$\pm$0.02\\
\phd PAH $10.8-14.5$ $\mu$m&~&4.45$\pm$0.13&~&0.68$\pm$0.02\\
\phd PAH $16.3-17.7$ $\mu$m&~&4.45$\pm$0.13&~&0.68$\pm$0.02\\
\phd 15 $\mu$m continuum\tablenotemark{c}&7.0$\pm$0.5&~&0.6.3$\pm$0.4&\\

\enddata

\tablenotetext{a}{Intensity units are 10$^{-17}$ erg cm$^{-2}$ s$^{-1}$ arcsec$^{-2}$}
\tablenotetext{b}{Upper limits are 3$\sigma$}
\tablenotetext{c}{Integrated over $14.5-15.5$ $\mu$m}

\end{deluxetable}

\begin{deluxetable}{lcc}
\tabletypesize{\scriptsize}
\tablecaption{NGC 891 Line Ratios\label{tbl-4}}
\tablewidth{0pt}
\tablehead{
\colhead{Position} & 
\colhead{[Ne$\,$III]/[Ne$\,$II]} &
\colhead{[S$\,$III]/[Ne$\,$II]} \\
}
\startdata

\phd East Halo &0.67$\pm$0.04\tablenotemark{a}&0.71$\pm$0.04\\
\phd West Halo&0.47$\pm$0.03&0.33$\pm$0.03\\
\enddata

\tablenotetext{a}{Errors are based on random noise in the data and the wavelength-dependent pointing uncertainty only.}
\end{deluxetable}

\begin{landscape}
\begin{deluxetable}{lcccccccc}
\tabletypesize{\scriptsize}
\tablecaption{NGC 5775 Infrared Line Intensities and PAH Equivalent Widths\label{tbl-5}}
\tablewidth{0pt}
\tablehead{
\colhead{} & \colhead{Disk 1} & \colhead{Disk 1} & \colhead{Disk 2} & \colhead{Disk 2} & \colhead{Disk 3} &\colhead{Disk 3} &\colhead{Halo} &\colhead{Halo} \\ 
\colhead{Line} & \colhead{Intensity\tablenotemark{a}} & \colhead{EW} & \colhead{Intensity} & \colhead{EW} & \colhead{Intensity} & \colhead{EW} & \colhead{Intensity} & \colhead{EW} \\
\colhead{} &
\colhead{} &
\colhead{($\mu$m)} &
\colhead{} &
\colhead{($\mu$m)} &
\colhead{} &
\colhead{($\mu$m)} &
\colhead{} &
\colhead{($\mu$m)} \\
}

\startdata

\phd [S$\,$IV]   10.51 $\mu$m&$<$3.7\tablenotemark{b}&~&$<$3.8&~&$<$3.9&~&$<$0.5&~\\
\phd [Ne$\,$II]  12.81 $\mu$m&96$\pm$7&~& 127$\pm$8&~&155$\pm$10&~&13.8$\pm$0.9&~\\
\phd [Ne$\,$III] 15.56 $\mu$m&17$\pm$2&~& 24$\pm$2&~&28$\pm$2&~&3.2$\pm$0.3&~\\
\phd [S$\,$III]  18.71 $\mu$m&31$\pm$2&~& 39$\pm$3&~&56$\pm$4&~&4.3$\pm$0.3&~\\
\phd H$_2$ {\it S}(1) $J=3-1$ 17.03 $\mu$m&29$\pm$2&~& 32$\pm$2&~&30$\pm$2&~&3.0$\pm$0.2&~\\
\phd PAH 11.2 $\mu$m&1045$\pm$72&1.05$\pm$0.03& 1006$\pm$70&1.03$\pm$0.03&972$\pm$67&1.05$\pm$0.03&84$\pm$6&1.12$\pm$0.04\\
\phd PAH 12.0 $\mu$m&60$\pm$7&0.05$\pm$0.005& 64$\pm$7&0.05$\pm$0.005&70$\pm$8&0.06$\pm$0.006&10$\pm$1&0.12$\pm$0.009\\
\phd PAH 12.7 $\mu$m&356$\pm$24&0.42$\pm$0.01& 338$\pm$23&0.40$\pm$0.01&368$\pm$25&0.51$\pm$0.01&23$\pm$2&0.41$\pm$0.01\\
\phd PAH 16.5 $\mu$m&55$\pm$4&0.11$\pm$0.007& 57$\pm$5&0.12$\pm$0.008&54$\pm$5&0.12$\pm$0.008&4.4$\pm$0.4&0.11$\pm$0.009\\
\phd PAH 17.4 $\mu$m&28$\pm$3&0.06$\pm$0.005& 23$\pm$3&0.04$\pm$0.005&21$\pm$3&0.04$\pm$0.006&3.9$\pm$0.4&0.10$\pm$0.009\\
\phd PAH $10.8-14.5$ $\mu$m&~&3.4$\pm$0.1&~&3.6$\pm$0.1&~&3.75$\pm$0.1&~&3.83$\pm$0.1\\
\phd PAH $16.3-17.7$ $\mu$m&~&0.68$\pm$0.02&~&0.78$\pm$0.02&~&0.58$\pm$0.01&~&0.71$\pm$0.02\\
\phd 15 $\mu$m continuum\tablenotemark{c}&427$\pm$28&~& 427$\pm$28&~&402$\pm$26&~&33$\pm$2&\\

\enddata

\tablenotetext{a}{Units are 10$^{-17}$ erg cm$^{-2}$ s$^{-1}$ arcsec$^{-2}$}
\tablenotetext{b}{Upper limits are 3$\sigma$}
\tablenotetext{c}{Integrated over $14.5-15.5$ $\mu$m}
\end{deluxetable}
\end{landscape}

\begin{deluxetable}{lcc}
\tabletypesize{\scriptsize}
\tablecaption{NGC 5775 Line Ratios\label{tbl-6}}
\tablewidth{0pt}
\tablehead{
\colhead{Position} & 
\colhead{[Ne$\,$III]/[Ne$\,$II]} &
\colhead{[S$\,$III]/[Ne$\,$II]} \\
}
\startdata

\phd Disk 1&0.18$\pm$0.02\tablenotemark{a}&0.32$\pm$0.01\\
\phd Disk 2&0.19$\pm$0.01&0.31$\pm$0.01\\
\phd Disk 3&0.18$\pm$0.01&0.36$\pm$0.01\\
\phd Halo&0.23$\pm$0.01&0.31$\pm$0.01\\

\enddata

\tablenotetext{a}{Errors are based on random noise in the data and the wavelength-dependent pointing uncertainty only.}
\end{deluxetable}

\begin{landscape}
\begin{deluxetable}{lcccccc}
\tabletypesize{\scriptsize}
\tablecaption{NGC 3044 Disk Infrared Line Intensities and PAH Equivalent Widths\label{tbl-7}}
\tablewidth{0pt}
\tablehead{
\colhead{} & \colhead{Disk 1} & \colhead{Disk 1} & \colhead{Disk 2} & \colhead{Disk 2} & \colhead{Disk 3} &\colhead{Disk 3} \\ 
\colhead{Line} & \colhead{Intensity\tablenotemark{a}} & \colhead{EW} & \colhead{Intensity} & \colhead{EW} & \colhead{Intensity} & \colhead{EW} \\
\colhead{} &
\colhead{} &
\colhead{($\mu$m)} &
\colhead{} &
\colhead{($\mu$m)} &
\colhead{} &
\colhead{($\mu$m)} \\
}

\startdata

\phd [S$\,$IV]  10.51 $\mu$m&$<$1.1&~&$<$2.4\tablenotemark{b}&~&$<$4.3&~\\
\phd [Ne$\,$II] 12.81 $\mu$m&30.8$\pm$2.1&~&82$\pm$5&~& 184$\pm$12&~\\
\phd [Ne$\,$III] 15.56 $\mu$m&18.5$\pm$1.2&~&61$\pm$4&~& 54$\pm$4&~\\
\phd [S$\,$III]  18.71 $\mu$m&11.9$\pm$0.8&~&27$\pm$2&~& 48$\pm$3&~\\
\phd H$_2$ {\it S}(1) $J=3-1$ 17.03 $\mu$m&11.2$\pm$0.8&~&12$\pm$1&~& 38$\pm$3&~\\
\phd PAH 11.2 $\mu$m&201$\pm$14&1.05$\pm$0.03&561$\pm$39&0.82$\pm$0.02& 1066$\pm$74&0.84$\pm$0.02\\
\phd PAH 12.0 $\mu$m&31$\pm$3&0.17$\pm$0.01&33$\pm$5&0.05$\pm 0.01$&77$\pm$8&0.05$\pm$0.01\\
\phd PAH 12.7 $\mu$m&82$\pm$6&0.68$\pm$0.01&215$\pm$15&0.46$\pm$0.01& 423$\pm$27&0.44$\pm$0.01\\
\phd PAH 16.5 $\mu$m&7.2$\pm$0.9&0.09$\pm$0.01&12$\pm$2&0.04$\pm$0.01& 73$\pm$6&0.12$\pm$0.01\\
\phd PAH 17.4 $\mu$m&4.6$\pm$0.8&0.06$\pm$0.01&15$\pm$2&0.06$\pm$0.01& 23$\pm$3&0.04$\pm$0.01\\
\phd PAH $10.8-14.5$ $\mu$m&~&3.2$\pm$0.1&~&2.8$\pm$0.1&~&3.1$\pm$0.1\\
\phd PAH $16.3-17.7$ $\mu$m&~&0.42$\pm$0.01&~&0.23$\pm$0.07&~&0.49$\pm$0.01\\
\phd 15 $\mu$m continuum\tablenotemark{c}&75.3$\pm$4.9&~&276$\pm$18&~&527$\pm$34&\\

\enddata

\tablenotetext{a}{Units are 10$^{-17}$ erg cm$^{-2}$ s$^{-1}$ arcsec$^{-2}$}
\tablenotetext{b}{Upper limits are 3$\sigma$}
\tablenotetext{c}{Integrated over $14.5-15.5$ $\mu$m}

\end{deluxetable}
\end{landscape}

\begin{deluxetable}{rcccc}
\tabletypesize{\scriptsize}
\tablecaption{NGC 3044 Halo Infrared Line Intensities and PAH Equivalent Widths\label{tbl-8}}
\tablewidth{0pt}

\tablehead{
\colhead{} & 
\colhead{North Halo} & \colhead{North Halo} & \colhead{South Halo} & \colhead{South Halo} \\
\colhead{Line} & \colhead{Intensity\tablenotemark{a}} & \colhead{EW} & \colhead{Intensity} & \colhead{EW} \\
\colhead{} &
\colhead{} &
\colhead{$\mu$m} &
\colhead{} &
\colhead{$\mu$m} \\
}
\startdata

\phd [S$\,$IV]  10.51 $\mu$m&$<$0.18&~&$<$0.18\tablenotemark{b}&~\\
\phd [Ne$\,$II] 12.81 $\mu$m&4.8$\pm$0.3&~&4.4$\pm$0.3&~\\
\phd [Ne$\,$III]15.56 $\mu$m&1.7$\pm$0.1&~&3.7$\pm$0.2&~\\
\phd [S$\,$III] 18.71 $\mu$m&2.4$\pm$0.2&~&1.23$\pm$0.09&~\\
\phd H$_2$ {\it S}(1) $J=3-1$ 17.03 $\mu$m&3.7$\pm$0.2&~&1.9$\pm$0.1&~\\
\phd PAH 11.2 $\mu$m& 36.9$\pm$2.6&1.15$\pm$0.04&33.5$\pm$2.3&1.28$\pm$0.04\\
\phd PAH 12.0 $\mu$m& 2.7$\pm$0.3&0.08$\pm$0.008&6.1$\pm$0.5&0.22$\pm$0.01\\
\phd PAH 12.7 $\mu$m& 9.6$\pm$0.7&0.40$\pm$0.01&12.5$\pm$0.9&0.57$\pm$0.02\\
\phd PAH 16.5 $\mu$m& 3.8$\pm$0.3&0.21$\pm$0.01&2.4$\pm$0.2&0.15$\pm$0.01\\
\phd PAH 17.4 $\mu$m& 1.2$\pm$0.2&0.07$\pm$0.01&2.5$\pm$0.2&0.17$\pm$0.01\\
\phd PAH $10.8-14.5$ $\mu$m&~&3.4$\pm$0.1&~&3.3$\pm$0.1\\
\phd PAH $16.3-17.7$ $\mu$m&~&0.74$\pm$0.02&~&0.68$\pm$0.02\\
\phd 15 $\mu$m continuum\tablenotemark{c}&15.1$\pm$1.0&~&12.6$\pm$0.8&\\
\enddata

\tablenotetext{a}{Units are 10$^{-17}$ erg cm$^{-2}$ s$^{-1}$ arcsec$^{-2}$}
\tablenotetext{b}{Upper limits are 3$\sigma$}
\tablenotetext{c}{Integrated over $14.5-15.5$ $\mu$m}
\end{deluxetable}

\begin{deluxetable}{lcc}
\tabletypesize{\scriptsize}
\tablecaption{NGC 3044 Line Ratios\label{tbl-9}}
\tablewidth{0pt}
\tablehead{
\colhead{Position} & 
\colhead{[Ne$\,$III]/[Ne$\,$II]} &
\colhead{[S$\,$III]/[Ne$\,$II]} \\
}
\startdata

Disk 1&0.60$\pm$0.02\tablenotemark{a}&0.39$\pm$0.01\\
Disk 2&0.74$\pm$0.02&0.32$\pm$0.01\\
Disk 3&0.29$\pm$0.01&0.26$\pm$0.01\\
North Halo&0.35$\pm$0.02&0.49$\pm$0.02\\
South Halo&0.84$\pm$0.03&0.28$\pm$0.01\\

\enddata
\tablenotetext{a}{Errors are based on random noise in the data and the wavelength-dependent pointing uncertainty only.}
\end{deluxetable}

\begin{deluxetable}{lccc}
\tabletypesize{\scriptsize}
\tablecaption{Equivalent Exponential Scale Heights for Gas and PAHs in NGC 891 and NGC 5775, Uncorrected for Extinction\label{tbl-10}}
\tablewidth{0pt}
\tablehead{
\colhead{Line} & 
\colhead{NGC 891 East Halo} &
\colhead{NGC 891 West Halo} &
\colhead{NGC 5775} \\
}
\startdata

\phd [Ne$\,$II] 12.81 $\mu$& $420 \pm 100$ & $400 \pm 50$ & $890 \pm 180$\\  
\phd [Ne$\,$III]15.56 $\mu$& $640 \pm 190$ & $540 \pm 80$ & $1000 \pm 260$\\  
\phd [S$\,$III] 18.71 $\mu$& $530 \pm 240$ & $470 \pm 40$ & $860 \pm 190$\\  
\phd H$_2$ {\it S}(1) $J=3-1$ 17.03 $\mu$m&$ 580 \pm 160$ & $550 \pm 80$ & $850 \pm 60$\\  
\phd PAH 11.2 $\mu$m& $470 \pm 60$ & $430 \pm 10$ & $790 \pm 50$\\  
\phd PAH 12.0 $\mu$m& $530 \pm 250$ & $510 \pm 80$ & $1060 \pm 150$\\  
\phd PAH 12.7 $\mu$m& $480 \pm 80$ & $440 \pm 40$ & $720 \pm 80$\\  
\phd PAH 16.5 $\mu$m& $460 \pm 40$ & $460 \pm 20$ & $780 \pm 60$\\  
\phd PAH 17.4 $\mu$m& $470 \pm 30$ & $500 \pm 50$ & $1080 \pm 220$\\  
\phd PAH 15 $\mu$m continuum& $440 \pm 100$ & $380 \pm 20$ & $780 \pm 40$\\

\enddata

\end{deluxetable}

\begin{deluxetable}{ccccccc}
\tabletypesize{\scriptsize}
\tablecaption{Derived physical parameters vs. height in the halo of NGC 891\tablenotemark{a}\label{tbl-11} }
\tablehead{  \colhead{Height} & \colhead{$\log U$} & \colhead{$\log T$} & \colhead{$\log T_{b}$} & \colhead{$\Phi_{H}$\tablenotemark{b}} & \colhead{$T$} & \colhead{$T_{b}$}  \\
(kpc) &  &  &  & (${\rm 10^{6}~ph~s^{-1}~cm^{-2}}$) & (K) & (K)}
\startdata
$z=0$\tablenotemark{c} & $-3.10 \pm 0.10$ & $4.56 \pm 0.03$  & $3.85 \pm 0.05$ & ($2.40 \pm 0.55$)$n_{H,-1}$ & $7130 \pm 820$ & $36,000 \pm 2000$ \\ 
$z=1$                           & $-3.48 \pm 0.06$ & $4.66 \pm 0.03$  & $3.86 \pm 0.03$ & ($1.00 \pm 0.14$)$n_{H,-1}$ & $7260 \pm 500$ & $45,800 \pm 3200$ \\ 
$z=2$                           & $-3.80 \pm 0.05$ & $4.76 \pm 0.03$  & $3.94 \pm 0.02$ & ($0.48 \pm 0.05$)$n_{H,-1}$ & $8600 \pm 300$ & $57,000 \pm 3300$ \\
\enddata
\tablenotetext{a}{As derived from matching [Ne$\,$III]/[Ne$\,$II], [S$\,$II]/H$\alpha$, [N$\,$II]/H$\alpha$, and [O$\,$III]/H$\alpha$. The difficulties of matching [O$\,$I/H$\alpha$, He$\,$I 5876/H$\alpha$, and [S$\,$III]/[Ne$\,$II], and implications thereof, are discussed in the text.} 
\tablenotetext{b}{All fluxes of hydrogen ionizing photons are scaled to the same density. Assuming a gas scaleheight of $h_{z}=2.3$ kpc and no change in gas filling factor would reduce the fluxes at $z=1$ kpc and $z=2$ kpc by a factor of 0.65 and 0.42 respectively.} 
\tablenotetext{c}{Emission line ratios come closest together at listed parameters, but there is no unique solution.} 
\label{partable}
\end{deluxetable}


\end{document}